\documentclass[prd,aps,nofootinbib,preprintnumbers,showpacs,showkeys]
{revtex4}
\usepackage{graphicx,epsf,amsfonts,amssymb,amsbsy}
\newcommand{\ds}{\displaystyle}
\newcommand{\vev}[1]{\langle#1\rangle}
\newcommand{\mat}{\left ( \begin{array}}
\newcommand{\emat}{\end{array} \right )}
\newcommand{\vect}{\left ( \begin{array}{c}}
\newcommand{\evect}{\end{array} \right )}

\preprint{HU-EP-05/19}

\begin{document}
\title{ \bf Mesons and diquarks in the color neutral 2SC phase of
dense cold quark matter }
\author{D.~Ebert}
\email{debert@physik.hu-berlin.de}
\affiliation{Institut f\"ur Physik,
Humboldt-Universit\"at zu Berlin, 12489 Berlin, Germany}
\author{K. G.~Klimenko}
\email{kklim@ihep.ru}
\affiliation{Institute
of High Energy Physics, 142281, Protvino, Moscow Region, Russia}
\author{V. L. Yudichev}
\email{yudichev@thsun1.jinr.ru}
\affiliation{Joint Institute for Nuclear Research, 141980, Dubna,
Moscow Region, Russia}

\begin{abstract}
The spectrum of  meson and diquark excitations of  dense color
neutral cold quark matter is investigated in the framework of a
2-flavored Nambu--Jona-Lasinio type model, including a quark $\mu$-
and color $\mu_8$ chemical potential. It was found out that in the
color superconducting (2SC) phase, i.e. at $\mu>\mu_c=342$ MeV,
$\mu_8$ aquires rather small values $\sim$ 10 MeV in order to ensure
the color neutrality. In this phase the $\pi$- and $\sigma$ meson
masses are evaluated around $\sim$ 330 MeV. The spectrum of scalar
diquarks in the color neutral 2SC phase consists of a heavy ($\rm
SU_c(2)$-singlet) resonance with mass $\sim$ 1100 MeV, four light
diquarks with mass $3|\mu_8|$, and one Nambu --Goldstone boson which
is in accordance with the Goldstone theorem. Moreover, in the 2SC
phase there are five light stable particles as well as a heavy
resonance in the spectrum of pseudo-scalar diquarks. In the color
symmetric phase, i.e. for $\mu <\mu_c$, a mass splitting of scalar
diquarks and antidiquarks is shown to arise if $\mu\ne 0$, contrary
to the case of $\mu = 0$, where the masses of scalar antidiquarks and
diquarks are degenerate at the value $\sim$~700 MeV. If the coupling
strength in the pseudo-scalar diquark channel is the same as in the
scalar diquark one (as for QCD-inspired NJL models), then in the
color symmetric phase pseudo-scalar diquarks are not allowed to exist
as stable particles.
\end{abstract}

% 11.30.Qc Spontaneous and radiative symmetry breaking
% 12.39.-x Phenomenological quark models
% 11.15.Ex Spontaneous breaking of gauge symmetries
% 12.38.Aw General properties of QCD
% 12.38.Mh Quark-gluon plasma
% 11.10.St Bound and unstable states; Bethe-Salpeter equations
% 12.38.-t Quantum chromodynamics
% 12.38.Lg Other nonperturbative calculations
% 26.60.+c Nuclear matter aspects of neutron stars
% 21.65.+f Nuclear matter
% 12.39.Fe
% 14.40.-n 

\maketitle
%\draft
%\large
%\maketitle

\section{Introduction}

Recent investigations, performed in the framework of perturbative
QCD, show that at low temperatures and asymptotically
high values of the quark chemical potential $\mu$ the dense baryonic
matter is a color superconductor \cite{son}. Evidently, at rather
small values of $\mu$ a more adequate description of this phenomenon
can be done with the help of different effective models, such as
Nambu -- Jona-Lasinio- (NJL) type field theories with four-fermionic
interaction \cite{njl}, etc. In such a way, on the basis of NJL-type
models with two quark flavors it was shown
that the color superconducting (2SC) phase might be yet present at
rather small values of $\mu\sim 350$ MeV, i.e. at baryon densities
only several times larger, than the density of ordinary nuclear
matter (see reviews \cite{alford,hs}). (For simplicity, throughout
the paper we will study quark matter, composed from two quark
flavors, i.e. up- and down quarks, only.) This is just the density of
compact star cores. So, color superconductivity, which is eventually
existing inside compact stars, might influence different observable
astrophysical processes, and thus deserves to be studied in more
details.

In the early NJL approach to color superconductivity
\cite{skp,bvy03}, the density of the color charge $Q_8$ was not
constraint to zero in the 2SC ground state
(the densities of other color charges $Q_i$, where $i=1,..,7$, are
zero in this phase \cite{bh}) leading to a nonvanishing
difference between the densities of red/green quarks and
blue ones. In this case, the mesonic and diquark excitations of dense
quark matter were considered in the framework of a simple
two-flavored NJL model with a single quark chemical potential $\mu$
\cite{bekvy,eky}. In particular, it was shown that in the color
asymmetric 2SC phase of this model does arise an abnormal number of
three Nambu -- Goldstone bosons (NG) instead of expected five
\footnote{Recall, that in the 2SC phase of the 2-flavored NJL model
the initial $\rm SU_c(3)$ color symmetry is spontaneously broken down
to the $\rm SU_c(2)$ one, so one might expect naively five massless
bosons. However, due to Lorentz noninvariance of the system, there
are indeed only three NG bosons.}, and $\pi$-mesons are stable
excitations of its ground state with masses $\sim 300$ MeV. Besides,
there are two light stable scalar diquark modes, whose masses are
proportional to $\vev{Q_8}$, as well as one heavy scalar diquark
resonance in the 2SC phase.

In reality, however, i.e. possibly in compact star cores or in
relativistic heavy ion experiments, there are several physical
constraints on the quark matter. The most evident one is its color
neutrality, which means a vanishing of a bulk $Q_i$ color charges
($i=1,...,8$). Indeed, since the lump of quark matter, which might be
created after heavy ion collisions, originated from color neutral
objects and is surrounded by a color neutral medium, it is expected
to be globally color neutral. To fulfill this requirement, usually
the local color neutrality constraint is imposed by introducing
several new chemical potentials, $\mu_3$, $\mu_8$, etc, into a NJL
model \cite{hs,mhuang,zhuang}. Otherwise, there will be produced a
chromo-electric field resulting in the flow of color charges, so a
homogeneous and color conducting quark medium with nonzero color
charge densities is not allowed to be a stable one \cite{mhuang}.
(Recall, there is no need to add new chemical potentials into QCD.
The point is that in the QCD 2SC ground state a nonzero value of the
eightth gluon field component might be generated. Effectively, it is
the $\mu_8$-chemical potential, so color neutrality is fulfilled
automatically in the QCD approach \cite{rebhan}.)

In the present paper, in contrast to our previous investigations
\cite{bekvy,eky}, we study the mesonic and diquark excitations of
color neutral quark matter \footnote{ Note that in compact star
cores one should consider the electrically neutral quark matter in
beta equilibrium, whereas in heavy-ion experiments the isospin and
strangeness are conserved quantities. For simplicity , these
additional constraints on the quark matter are ignored in the present
consideration.} in the framework of a simple two flavor NJL model at
zero temperature. We consider both the case of rather
small values of the quark chemical potential $\mu$ (the $SU_c(3)$
color symmetric (normal) phase), and the case of $\mu$-values,
corresponding to the 2SC phase of the model. In addition, the
properties of pseudo-scalar diquarks are also included into the
consideration.

The paper is organized as follows. In Section II the thermodynamic
potential as well as the effective action of the NJL model, extended
by an additional color $\mu_8$-chemical potential term, is obtained
in the one quark loop approximation. Further, in Section III, the gap
equations and the phase structure of the model is investigated under
the color neutrality constraint. Here the values of $\mu_8$ are
obtained at which the 2SC phase is a color neutral one. Then, in the
Sections IV, V, and VI, the masses of the $\pi$- and $\sigma$ mesons,
scalar diquarks, and pseudo-scalar diquarks are considered,
respectively. Finally, in the Appendix, the influence of the mixing
between $\sigma$-meson and scalar diquarks on the $\sigma$-mass is
discussed.

\section{The model and its effective action}
\label{II}

In the original version of the NJL model \cite{njl} the
four-fermionic interaction of a proton $p$ and neutron $n$ doublet
was considered, and the principle of dynamical chiral symmetry
breaking was demonstrated. Later, the $(p,n)$-doublet was replaced
by a doublet of colored up $u$ and down $d$ quarks,
in order to describe phenomenologically the physics of
light mesons \cite{ebvol,volk,hatsuda,volkyud},
diquarks \cite{ebka,vog}, as well as meson-baryon interactions
\cite{ebjur,reinh}. In this sense the NJL model may be thought of
as an effective theory for low energy QCD. \footnote{Indeed,
consider two-flavor QCD, symmetric under the color group $\rm
SU_c(3)$. By integrating in the generating functional of QCD over
gluons and further ``approximating'' the nonperturbative gluon
propagator by a $\delta-$function, one arrives at an effective local
chiral four-quark interaction of the NJL type describing low-energy
hadron physics. Moreover, by performing a Fierz transformation of the
interaction terms, it possible to obtain a NJL-type Lagrangian
describing the interaction of quarks in the scalar and
pseudo-scalar $(\bar q q)$- as well as scalar and pseudo-scalar
diquark $(qq)$ channels (see, e.g., the Lagrangian (\ref{1}) below).}
(Of course, it is necessary to remember that in the NJL model, in
contrast with QCD, quarks are not confined in the hardronic phase,
which is a shortcoming of the model.)
At present time the phenomenon of dynamical (chiral) symmetry
breaking is one of the cornerstones of modern physics. So, this
effect was studied in the framework of NJL-type models under the
influence of external magnetic fields \cite{mir}, in curved
space-times \cite{odin}, in spaces with nontrivial topology
\cite{incera}, etc. Formally, as it was mentioned above, quarks are
presented in the mass spectrum of the model. So it is very suitable
for the description of normal hot and/or dense quark matter
\cite{hatsuda,asakawa,ebert,klim} in which, as it is known, quarks
are not confined. NJL-type models still remain a simple but useful
instrument for studying color superconducting quark matter at
moderate densities \cite{alford,hs,skp,bvy03}, where analytical and
lattice computations in the framework of QCD are impossible.

We start from the following 2-flavor NJL Lagrangian, called for the
description of interactions in the quark-antiquark, scalar diquark-
as well as pseudo-scalar diquark channels at low and moderate
energies and baryon densities (the consideration is performed in
Minkovski space-time notation):
\begin{eqnarray}
&&  L_q=\bar q\Big [\gamma^\nu i\partial_\nu-
m_0+\mu\gamma^0\Big ]q+ G\Big\{(\bar qq)^2+
(\bar qi\gamma^5\vec\tau q)^2\Big\}+\nonumber\\
&+&\!\!\!\sum_{A=2,5,7}\Big\{H_s
[\bar q^Ci\gamma^5\tau_2\lambda_{A}q]
[\bar qi\gamma^5\tau_2\lambda_{A} q^C]
+H_p [\bar q^C\tau_2\lambda_{A}q]
[\bar q\tau_2\lambda_{A} q^C]\Big\},
  \label{1}
\end{eqnarray}
where $\mu >0$ is the quark chemical potential, the quark
field $q\equiv q_{i\alpha}$ is a flavor doublet and color triplet as
well as a four-component Dirac spinor, where $i=1,2$ (or $i=u,d$) and
$\alpha=1,2,3$ (or $\alpha=r,g,b$). $q^C=C\bar q^t$, $\bar q^C=q^t C$
are charge-conjugated spinors, and $C=i\gamma^2\gamma^0$ is the
charge conjugation matrix (symbol $t$ denotes the transposition
operation). It is supposed that up and down quarks have equal current
(bare) mass $m_0$. Furthermore, we use the notations  $\tau_a$ for
Pauli matrices, and $\lambda_A$ for antisymmetric Gell-Mann matrices
in flavor and color space, respectively. Clearly, the Lagrangian
$L_q$ is invariant under transformations from the color $\rm
SU_c(3)$- as well as baryon $\rm U_B(1)$ groups. In addition, at
$m_0=0$ this Lagrangian is symmetric under the chiral $\rm
SU(2)_L\times SU(2)_R$ group (chiral transformations act on the
flavor indices of quark fields only). Moreover, since $Q=I_3+B/2$,
where $I_3=\tau_3/2$ is the generator of the third isospin component,
$Q$ is the generator of the electric charge, and $B$ is the baryon
charge generator, our system is symmetric under the electromagnetic
group $\rm U_Q(1)$ as well. If the Lagrangian (\ref{1}) is
obtained from the QCD one-gluon exchange approximation, then
$G:H_s:H_p=4:3:3$. However, in the present consideration we will
not fix relations between coupling constants, so they are treated as
free parameters. It is necessary to note also that at $\mu =0$ the
Lagrangian (\ref{1}) is invariant under the charge conjugation
symmetry ($q\to q^C\equiv C\bar q^t$, $\bar q\to \bar q^C\equiv q^t
C$) that is, however, spoiled by the chemical potential term.

Furtheron, the temperature is chosen to be zero throughout the paper.
In this case there is a critical value $\tilde\mu_c$ of the chemical
potential, such that at $\mu<\tilde\mu_c$ the color symmetric
(normal) phase is realized in the system (evidently, the ground state
of this phase is a color singlet). However, at $\mu>\tilde\mu_c$ one
obtains a color superconducting (2SC) phase in which $\rm SU_c(3)$ is
spontaneously broken down to $\rm SU_c(2)$. Only two quark colors,
say red and green, participate in the gap formation in the 2SC phase,
the blue quarks stay gapless. So the densities of red/green quarks,
$n_{r,g}$, are equal in this phase, however the density $n_b$ of blue
quarks is not equal to $n_{r,g}$, i.e. local color neutrality is
broken. \footnote{\label{Q_3} Thus, $\vev{Q_8}\ne 0$. However, the
color charge $Q_3=\bar q\gamma^0T_3q$, where $T_3=$diag$(1,-1,0)$ is
the matrix in the color space, vanishes in the 2SC phase. Other color
charges $Q_i$ $(i\ne 8)$ are also zero in this phase \cite{bh}.}
To restore local color neutrality in the 2SC phase of the model
(\ref{1}), usually an additional chemical potential term $\mu_8Q_8$
is introduced into the considerations \cite{hs}, where $Q_8=\bar
q\gamma^0T_8q$, and $T_8=$diag$(1,1,-2)=\sqrt{3}\lambda_8$ is the
matrix in the color space. Hence,
\begin{eqnarray}
L_q\to L=L_q+\mu_8Q_8.
\label{3}
\end{eqnarray}
If $\mu_8$ is an independent model parameter, then Lagrangian
(\ref{3}) is symmetric under the color $\rm SU_c(2)\times
U_{\lambda_8}(1)$ group. However, if local color neutrality is
imposed, then the chemical potential $\mu_8$ is not an independent
parameter. Its value must be chosen in such a way that the ground
state expectation value $\vev{Q_8}$ is identically equal to zero.
Hence, $\mu_8$ depends on $\mu$ etc. For example, in the color
symmetric phase, i.e. at $\mu<\mu_c$ (it will be shown in section
\ref{III} that in general case $\mu_c\ne $ $\tilde\mu_c$), where
$\vev{Q_8}\equiv 0$ even in the theory (\ref{1}), we have to put
$\mu_8\equiv 0$. However $\mu_8$ has a nontrivial
$\mu$-dependence at $\mu>\mu_c$ in order to supply the zero value of
$\vev{Q_8}$. As a consequence, we see that the color symmetry group
of the model (\ref{3}) depends on $\mu$: at $\mu<\mu_c$ it is
$\rm SU_c(3)$, whereas at $\mu>\mu_c$ it is the $\rm SU_c(2)\times
U_{\lambda_8}(1)$ subgroup of $\rm SU_c(3)$ which is just the color
symmetry group of the term $\mu_8Q_8$.

In the present paper we are going to study both the ground state
properties of the system with Lagrangian $L$, and its mass spectrum
in the quark-, meson- and diquark sectors. So, we have to obtain the
thermodynamic potential $\Omega$ as well as the effective action of
the model up to a second order in the bosonic degrees of freedom.
To begin with, let us introduce the linearized version of Lagrangian
$L$ that contains auxiliary bosonic fields:
\begin{eqnarray}
\tilde L\ds &=&\bar q\Big [\gamma^\nu i\partial_\nu +\mu\gamma^0+
\mu_8T_8 -\sigma - m_0 -i\gamma^5\pi_a\tau_a\Big ]q
 -\frac{1}{4G}\Big [\sigma\sigma+\pi_a\pi_a\Big ]-
 \frac1{4H_s}\Delta^{s*}_{A}\Delta^s_{A}-
 \frac1{4H_p}\Delta^{p*}_{A'}\Delta^p_{A'}-
 \nonumber\\ &-&
\frac{\Delta^{s*}_{A}}{2}[\bar q^Ci\gamma^5\tau_2\lambda_{A} q]
-\frac{\Delta^s_{A}}{2}[\bar q i\gamma^5\tau_2\lambda_{A}q^C]
-\frac{\Delta^{p*}_{A'}}{2}[\bar q^C\tau_2\lambda_{A'} q]
-\frac{\Delta^p_{A'}}{2}[\bar q\tau_2\lambda_{A'} q^C],
\label{8}
\end{eqnarray}
where as well as in the following the summation over  repeated
indices $a=1,2,3$ and $A,A'=2,5,7$ is implied.
 Lagrangians $L$ and $\tilde L$ are equivalent on the
equations of motion for bosonic fields, from which it follows
that
\begin{eqnarray}
\sigma (x)=-2G(\bar qq),~~~\Delta^s_{A}(x)\!\!&=&\!\!-2H_s(\bar
q^Ci\gamma^5\tau_2\lambda_{A}q),~~~
\Delta^{s*}_{A}(x)=-2H_s(\bar qi\gamma^5\tau_2\lambda_{A} q^C),
\nonumber\\\pi_a(x)=-2G(\bar qi\gamma^5\tau_a q),~~~
\Delta^p_{A'}(x)\!\!&=&\!\!-2H_p(\bar q^C\tau_2\lambda_{A'}q),~~~
\Delta^{p*}_{A'}(x)=-2H_p(\bar q \tau_2\lambda_{A'} q^C).
\label{9}
\end{eqnarray}
One can easily see from (\ref{9}) that mesonic fields $\sigma,\pi_a$
are real quantities, i.e. $(\sigma(x))^\dagger=\sigma(x),~~
(\pi_a(x))^\dagger=\pi_a(x)$ (the superscript symbol $\dagger$
denotes the hermitian conjugation), but all diquark fields
$\Delta^{s,p}_{A}$ are complex ones, so
\[
(\Delta^s_{A}(x))^\dagger=\Delta^{s*}_{A}(x),~~~~~
(\Delta^p_{A'}(x))^\dagger=\Delta^{p*}_{A'}(x).
\]
Moreover, $\Delta^s_{A}$ and $\Delta^p_{A'}$ are scalars and
pseudo-scalars, correspondingly. Clearly, the real $\sigma$ and
$\pi_a$ fields are color singlets, all scalar diquarks
$\Delta^s_{A}$ form a color antitriplet $\bar 3_c$ of the
$\rm SU_c(3)$ group. The same is true for pseudo-scalar diquarks
$\Delta^p_{A'}$ which are also the components of an $\bar
3_c$-multiplet of the color group. Evidently, in the $\rm
SU_c(3)$-color
symmetric phase (at $\mu<\mu_c$) all diquark fields must have zero
ground state expectation values, i.~e. $\vev{\Delta^s_{A}}= 0$ and
$\vev{\Delta^p_{A'}}= 0$. Otherwise, we have an indication
that the ground state of the system is no more an
$\rm SU_c(3)$-invariant one.

Lagrangian (\ref{8}) provides us with a common
footing for obtaining both the thermodynamic potential (TDP) and the
mass spectrum for bosonic excitations. Indeed, in the one-fermion
loop approximation, the effective action for the boson fields is
expressed through the path integral over quark fields:
$$
\exp(i {\cal S}_{\rm {eff}}(\sigma,\pi_a,\Delta^{s,p}_{A},
\Delta^{s,p*}_{A'}))=
  N'\int[d\bar q][dq]\exp\Bigl(i\int\tilde L\,d^4 x\Bigr),
$$
where
\begin{equation}
{\cal S}_{\rm {eff}}
(\sigma,\pi_a,\Delta^{s,p}_{A},\Delta^{s,p*}_{A'})
=-\int d^4x\left [\frac{\sigma^2+\pi^2_a}{4G}+
\frac{\Delta^s_{A}\Delta^{s*}_{A}}{4H_s}+
\frac{\Delta^p_{A'}\Delta^{p*}_{A'}}{4H_p}
\right ]+\tilde {\cal S}_{\rm {eff}},
\label{10}
\end{equation}
$N'$ is a normalization constant.
The quark contribution term $\tilde {\cal S}_{\rm {eff}}$ is here
given by:
\begin{equation}
\exp(i\tilde {\cal S}_{\rm {eff}})=N'\int [d\bar
q][dq]\exp\Bigl(\frac{i}{2}\int\Big [\bar
qD^+q+\bar q^CD^-q^C-\bar qK q^C-\bar
q^CK^{*}q\Big ]d^4 x\Bigr).
\label{11}
\end{equation}
In (\ref{11}) we have used the following notations
\begin{eqnarray}
&&D^+=i\gamma^\nu\partial_\nu- m_0+\hat\mu\gamma^0-\Sigma,~~~~~~~
D^-=i\gamma^\nu\partial_\nu- m_0-\hat\mu\gamma^0-\Sigma^t,
~~~~~~~ \Sigma=\sigma+ i\gamma^5\pi_a\tau_a,\nonumber\\
&&\Sigma^t=\sigma+ i\gamma^5\pi_a\tau^t_a,~~~~~
K^*=(\Delta^{p*}_{A}+i\Delta^{s*}_{A}\gamma^5)\tau_2\lambda_{A},
\qquad
K=(\Delta^p_{A}+i\Delta^s_{A}\gamma^5)\tau_2\lambda_{A},
\label{12}
\end{eqnarray}
where $D^\pm$ are nontrivial operators in the coordinate, spinor,
color and flavor spaces.\footnote{In order to bring the quark sector
of the Lagrangian (\ref{8}) to the expression, given in the square
brackets of (\ref{11}), we have also used the following well-known
relations:
$\partial_\nu^t=-\partial_\nu$, $C\gamma^\nu C^{-1}=-(\gamma^\nu)^t$,
$C\gamma^5C^{-1}=(\gamma^5)^t=\gamma^5$,
$\tau^2\vec\tau\tau^2=-(\vec\tau)^t$,
$\tau^2=\left (\begin{array}{cc}
0~, & -i\\
i~, &0
\end{array}\right )$.}
In the framework of the Nambu-Gorkov formalism, where quarks are
composed into a bispinor $\Psi =\left({q\atop q^C}\right)$, it is
possible to integrate in (\ref{11}) over quark fields and obtain
\begin{equation}
\tilde {\cal S}_{\rm
{eff}}(\sigma,\pi_a,\Delta^{s,p}_{A},\Delta^{s,p*}_{A'})
= \frac 1{2i}{\rm Tr}_{\{NGsfcx\}}\ln \left (\begin{array}{cc}
D^+~, & -K\\
-K ^*, &D^-
\end{array}\right )\equiv \frac 1{2i}{\rm Tr}_{\{NGsfcx\}}\ln Z.
\label{14}
\end{equation}
Besides of an evident trace in the two-dimensional Nambu-Gorkov space
(NG), the Tr-operation in (\ref{14}) stands for
calculating  the trace in spinor- ($s$), flavor- ($f$), color-
($c$) as well as four-dimensional coordinate- ($x$) spaces,
correspondingly.

Starting from (\ref{10})-(\ref{14}), it is possible to define the
thermodynamic potential $\Omega (\sigma,\pi_a, \Delta^{s,p}_{A},
\Delta^{s,p*}_{A'})$ of the model:
\begin{equation}
{\cal S}_{\rm {eff}}~\bigg
|_{~\sigma,\pi_a,\Delta^{s,p}_{A},\Delta^{s,p*}_{A'}=\rm
{const}}=-\Omega (\sigma,\pi_a,\Delta^{s,p}_{A},
\Delta^{s,p*}_{A'})\int d^4x,
\label{15}
\end{equation}
where, in the spirit of the mean-field approximation, all boson
fields are supposed to be $x$-independent.
It is well-known that ground state expectation values
$\vev{\sigma(x)}\equiv\sigma^o,~\vev{\pi_a(x)}\equiv\pi_a^o,~\vev
{\Delta^{s,p}_{A}(x)}\equiv\Delta^{s,po}_{A},~
\vev{\Delta^{s,p*}_{A'}(x)}\equiv\Delta^{s,p*o}_{A'}$ are
coordinates of the global minimum
point of the thermodynamic potential $\Omega$, i.e. they
form a solution of the gap equations (evidently, in our
approach all ground state expectation values do not depend on
coordinates $x$):
\begin{eqnarray}
\frac{\partial \Omega}{\partial\pi_a}=0,~~~~~
\frac{\partial \Omega}{\partial\sigma}=0,~~~~~
\frac{\partial \Omega}{\partial\Delta^{s,p}_{A}}=0,~~~~~
\frac{\partial \Omega}{\partial\Delta^{s,p*}_{A'}}=0.
\label{16}
\end{eqnarray}

Let us make the following shift of bosonic fields in ${\cal S}_{\rm
{eff}}$: $\sigma(x)\to\sigma(x)+\sigma^o$,
$\pi_a(x)\to\pi_a(x)+\pi_a^o$, $\Delta^{s,p*}_{A}(x)
\to\Delta^{s,p*}_{A}(x) +\Delta^{s,p*o}_{A}$, $\Delta^{s,p}_{A}(x)
\to\Delta^{s,p}_{A}(x)+\Delta^{s,p o}_{A}$, where
$\sigma^o,\pi^o_a,\Delta^{s,po}_{A},
\Delta^{s,p*o}_{A}$ have no coordinate dependency. In this case the
matrix $Z$ from (\ref{14}) is transformed in the following way:
\begin{equation}
Z\rightarrow\left (\begin{array}{cc}
D^+_o~, & -K_o\\
 -K_o^*~, & D^-_o
\end{array}\right )-\left (\begin{array}{cc}
\Sigma~, & K\\
 K^*~, & \Sigma^t
\end{array}\right )\equiv S_0^{-1}-\left (\begin{array}{cc}
\Sigma~, & K\\
 K^*~, & \Sigma^t
\end{array}\right ),
\label{17}
\end{equation}
where $S_0$ is the quark propagator matrix in the Nambu-Gorkov
representation, and
\[
(K_o, K^*_o, D^\pm_o,\Sigma_o,\Sigma^t_o)= (K, K^*, D^\pm,
\Sigma,\Sigma^t)~\bigg |_{~\sigma =\sigma^o, \pi_a =\pi_a^o, ...}
\]
Then, expanding the expression (\ref{10}) up to a second order over
the meson and diquark fields, we have
\begin{equation}
 {\cal S}_{\rm
 {eff}}(\sigma,\pi_a,\Delta^{s,p}_{A},\Delta^{s,p*}_{A'})=
 {\cal S}_{\rm {eff}}^{(0)} +
 {\cal S}_{\rm
 {eff}}^{(2)}(\sigma,\pi_a,\Delta^{s,p}_{A},\Delta^{s,p*}_{A'})
 +\cdots,
  \label{18}
\end{equation}
where (due to the gap equations, the term linear over meson and
diquark
fields is absent in (\ref{18}))
\begin{eqnarray}
 &&{\cal S}_{\rm {eff}}^{(0)}=-\int
 d^4x\left[\frac{\sigma^o\sigma^o+\pi^o_a\pi^o_a}{4G}+
\frac{\Delta^{so}_{A}\Delta^{s*o}_{A}}{4H_s}+
\frac{\Delta^{po}_{A'}\Delta^{p*o}_{A'}}{4H_p}\right]-\frac i2{\rm
Tr}_{\{NGscfx\}}\ln\left (S_0^{-1}\right )\equiv\nonumber\\
&&~~~~~~~~~~~~~~~~~~~~~~~~~~~~~
\equiv-\Omega (\sigma^o,\pi^o_a,\Delta^{s,po}_{A},
 \Delta^{s,p*o}_{A'})\int d^4x,
  \label{19}
\end{eqnarray}
\begin{eqnarray}
 {\cal S}^{(2)}_{\rm
 {eff}}(\sigma,\pi_a,\Delta^{s,p}_{A},\Delta^{s,p*}_{A'})
 \!\!\!\!&&\!\!=
 -\int d^4x\left[\frac{\sigma^2+\pi^2_a}{4G}+
\frac{\Delta^s_{A}\Delta^{s*}_{A}}{4H_s}+
\frac{\Delta^p_{A'}\Delta^{p*}_{A'}}{4H_p}\right]+\nonumber\\
&&~~~~~~~~~~~~~~~~~~~~~~+\frac i4{\rm
Tr}_{\{NGscfx\}}
\left\{S_0\left (\begin{array}{cc}
\Sigma~, & K\\
 K^*~, & \Sigma^t
\end{array}\right )S_0\left (\begin{array}{cc}
\Sigma~, & K\\
 K^*~, & \Sigma^t
\end{array}\right )\right\}.
  \label{21}
\end{eqnarray}
In the following on the basis of the effective action ${\cal S}_{\rm
{eff}}^{(2)}$ we will study the spectrum of meson/diquark excitations
in the color superconducting phase of the NJL model under
consideration. So, it is convenient to present the effective action
(\ref{21}) in the following form:
\begin{eqnarray}
{\cal S}^{(2)}_{\rm {eff}}={\cal S}^{(2)}_{\rm mesons}+{\cal
S}^{(2)}_{\rm diquarks}+{\cal S}^{(2)}_{\rm mixed},
\label{2.21}
\end{eqnarray}
where
\begin{eqnarray}
\label{2.22}
 {\cal S}^{(2)}_{\rm mesons} \!\!\!\!&&\!\!=
 -\int d^4x\frac{\sigma^2+\pi^2_a}{4G}+
\frac i4{\rm Tr}_{scfx}
\left\{S_{11}\Sigma S_{11}\Sigma +2S_{12}\Sigma^tS_{21}\Sigma +
S_{22}\Sigma^t S_{22}\Sigma^t\right\},\\
\label{2.23}
 {\cal S}^{(2)}_{\rm diquarks} \!\!\!\!&&\!\!=
 -\int d^4x\left[\frac{\Delta^s_{A}\Delta^{s*}_{A}}{4H_s}+
\frac{\Delta^p_{A'}\Delta^{p*}_{A'}}{4H_p}\right]
+\frac i4{\rm Tr}_{scfx}
\left\{S_{12}K^*S_{12}K^* +2S_{11}KS_{22}K^* +
S_{21}K S_{21}K\right\},\\
 {\cal S}^{(2)}_{\rm mixed} \!\!\!\!&&\!\!=
\frac i2{\rm Tr}_{scfx}
\left\{S_{11}\Sigma S_{12}K^* +S_{21}\Sigma S_{11}K +
S_{12}\Sigma^t S_{22}K^*+S_{21}KS_{22}\Sigma^t\right\},
 \label{2.24}
\end{eqnarray}
and $S_{ij}$ are the matrix elements of the quark propagator matrix
$S_0$, given in (\ref{17}).

\section{Gap equations and color neutrality condition }
\label{III}

Let us for a moment assume that in (\ref{3}) the chemical potential
$\mu_8$ is an independent parameter ($\ne 0$). Then, Lagrangian $L$
is invariant under the color $\rm SU_c(2)\times U_{\lambda_8}(1)$
symmetry group. Recall, that the phase structure of any theory is
defined by a competition of its order parameters. In our case, the
order parameters (ground state expectation values), $\vev{\sigma(x)},
~\vev{\pi_a(x)}, ~\vev {\Delta^{s}_{A}(x)}, ~\vev{\Delta^{p}_{A'}
(x)}$, are obtained from a solution of the gap equations (see the
previous section). Since the consideration of the model (\ref{3})
with a total set of order parameters is a very hard task, we
shall assume that parity is conserved, i.e. $\vev{\pi_a(x)}=0,~
\vev{\Delta^{p}_{A'}(x)}=0$ (in the section \ref{VIA} some
arguments are presented, however, that at sufficiently high values of
pseudo-scalar coupling $H_p$ parity might be spontaneously broken
down), thus having to deal only with $\vev{\sigma(x)}$ and
$\vev{\Delta^{s}_{A} (x)}$. In this case three different phases
might exist in the model  (\ref{3}): {\bf i)} In the first one,
the normal phase, $\vev {\Delta^{s}_{A}(x)}=0$ for  all $A=2,5,7$. In
this phase the initial color symmetry remains intact. {\bf ii)} The
second one is a well-known 2SC phase with $\vev{\Delta^{s}
_{2}(x)}\ne 0$ and $\vev{\Delta^{s}_{5,7}(x)} =0$. The ground state
of this phase is invariant under $\rm SU_c(2)$-color symmetry. {\bf
iii)} Finally, there might exist a phase with $\vev{\Delta
^{s}_{2}(x)} \ne 0$, $\vev{\Delta^{s}_{5}(x)} \ne 0$, and
$\vev{\Delta^{s}_{7}(x)}=0$. (Note, the two phases {\bf ii)} and {\bf
iii)} are not unitarily equivalent, since there are no color
transformations from $\rm SU_c(2)\times U_{\lambda_8}(1)$ that
connects the corresponding ground state expectation values.)
However, since color neutrality cannot be achieved in the ground
state of the form {\bf iii)} (see \cite{bh}), throughout of our paper
only two order parameters, $\vev{\sigma(x)}$ and $\vev{\Delta^{s}_{2}
(x)}\equiv \Delta$, will be taken into account whereas other ones
will be supposed to have zero expectation values, i.e.
$\vev{\pi_a(x)}=0,~\vev{\Delta^{p}_{A'}(x)}=0$, $\vev{\Delta^{s}
_{5,7}(x)} =0$. So, below only the competition between the normal
phase ( $\Delta =0$) and the 2SC one ( $\Delta\ne 0$) will be
considered. As a result, one may deal with a thermodynamic potential
$\Omega$ which depends on two variables $\Delta,\vev{\sigma}$
or, equivalently, $\Delta, m\equiv m_0+\vev{\sigma}$, only. Then, the
expression for the thermodynamic potential $\Omega$ can be calculated
with the help of (\ref{19}) (see also, e.g., \cite{zhuang}):
\begin{eqnarray}
&&\Omega(m,\Delta)=
\frac{(m-m_0)^2}{4G}+\frac{|\Delta|^2}{4H_s}-
4\sum_{\pm}\int\frac{d^3q}{(2\pi)^3}|E_{\Delta}^\pm|-
2\sum_{\pm}\int\frac{d^3q}{(2\pi)^3}|\breve E^\pm|,
\label{32}
\end{eqnarray}
where $E_\Delta^\pm=\sqrt{(E^\pm)^2+|\Delta|^2}$,
$E^\pm=E\pm\bar\mu$,
$\breve E^\pm=E\pm\breve\mu$, $\bar\mu=\mu+\mu_8$,
 $\breve\mu =\mu-2\mu_8$, $E=\sqrt{\vec q^2+m^2}$. Since the
 integrals in the right hand side of (\ref{32}) are
ultraviolet divergent, we regularize them and the other divergent
integrals below by implementing a three-dimensional cutoff $\Lambda$.
The resulting gap equations  look like:
\begin{eqnarray}
&&\frac{\Delta}{4H_s}=4i\Delta\int\frac{d^4q}{(2\pi)^4}
\Big\{\frac{1}{q_0^2-(E_\Delta^+)^2}+
\frac{1}{q_0^2-(E_\Delta^-)^2}\Big\}=2\Delta\int
\frac{d^3q}{(2\pi)^3}
\Big\{\frac{1}{E_\Delta^+}+
\frac{1}{E_\Delta^-}\Big\},
\label{33}\\
&&\frac{m-m_0}{2G}=4im\sum_{\pm}\int\frac{d^4q}{(2\pi)^4}
\frac{1}{E}\Big\{\frac{\breve E^\pm}{q_0^2-(\breve
E^\pm)^2}\Big\}+8im\sum_{\pm}\int\frac{d^4q}{(2\pi)^4}\frac{1}
{E}\Big\{\frac{E^\pm}{q_0^2-(E_\Delta^\pm)^2}\Big\}.
\label{34}
\end{eqnarray}
In (\ref{33}), (\ref{34}) as well as in other expressions, containing
four-dimensional momentum integrals, $q_0$ is shorthand for
$q_0+i\varepsilon\cdot {\rm sign}(q_0)$, where $\varepsilon\to 0_+$.
This prescription correctly implements the roles of $\mu$, $\mu_8$ as
chemical potentials and preserves the causality of the theory (see,
e.g. \cite{chodos}).

Now, let us impose the local color neutrality requirement on the
ground state of the model. It means that the quantity $\mu_8$ takes
such values that the density of the 8-color charge $\vev{Q_8}\equiv
-\partial\Omega/\partial\mu_8$ equals zero for arbitrary fixed values
of other model parameters. So, we have the following local color
neutrality constraint:
\begin{eqnarray}
&&\vev{Q_8}= -\frac{\partial\Omega}{\partial\mu_8}\equiv
4\int\frac{d^3q}{(2\pi)^3}
\Big\{\frac{E^+}{E_\Delta^+}- \frac{E^-}{E_\Delta^-}- \frac{\breve
E^+}{|\breve E^+|}+\frac{\breve E^-}{|\breve E^-|}\Big\}=0.
\label{35}
\end{eqnarray}
The system of eqs.\ (\ref{33})-(\ref{35}) has two different
solutions. As we have already discussed after (\ref{3}), the first
one (with $\Delta=0$, $\mu_8=0$) corresponds to the $\rm SU(3)
_c$-symmetric phase of the model (normal phase), the second one (with
$\Delta\ne 0$ and $\mu_8\ne 0$) to the 2SC phase.
As usual, solutions of these equations give  local extrema of the
thermodynamic potential $\Omega (m,\Delta)$ (\ref{32}),
so one should also check which of them corresponds to the absolute
minimum of $\Omega$. Having found the solution corresponding to the
stable state of quark matter (the absolute minimum of $\Omega$),
we obtained the behavior of the gaps $m$, $\Delta$ as well as the
$\mu_8$ vs. the quark chemical potential $\mu$ (see Figs~1,2). (Note,
in all numerical calculations of the paper we use the parameter set
\begin{eqnarray}
G=5.86~{\rm GeV}^{-2},~~~~~\Lambda=618~{\rm MeV},~~~~~m_0=5.67~ {\rm
MeV},~~~~~H_s=3G/4
\label{350}
\end{eqnarray}
that leads in the framework of the NJL model to the well-known vacuum
phenomenological values of the pion weak-decay constant
$F_\pi=92.4$~MeV, pion mass $M_\pi=140$~MeV, and chiral quark
condensate $\vev{\bar qq}=-(245~\mbox{MeV})^3$. The
relation between $H_s$ and $G$ in (\ref{350}) is induced, e.g., by
the structure of QCD four-quark vertices in the one-gluon exchange
approximation.) The region $\mu<\mu_c = 342$~MeV is the domain of
$\rm SU_c(3)$-color symmetric quark matter because $\Omega$  in this
case is minimized by $m\ne 0$ and $\Delta =0$, $\mu_8=0$. For $\mu>
\mu_c$, the solution with $m\ne 0$, $\Delta\ne 0$ and $\mu_8\ne 0$,
corresponding to the 2SC phase, gives the global minimum  of
$\Omega$, and thereby the color superconducting phase is favored.
The transition between these two phases is of the first-order, which
is characterized by a discontinuity in the behavior of $m$ and
$\Delta$ vs. $\mu$ (see Fig.~1). Finally, remark that the
critical value $\tilde\mu_c$ of the quark chemical potential was
calculated in the framework of the model (\ref{1}) as well. In terms
of the parameter set (\ref{350}) we have $\tilde\mu_c = 350$~MeV
\cite{bekvy,eky}, i.e. in the color neutral quark matter the
transition to the 2SC phase  occurred at slightly less values of the
quark chemical potentials.

Similarly to \cite{huang}, it is possible to find the
following expressions for the matrix elements $S_{ij}$ of the quark
propagator matrix $S_0$:
\begin{eqnarray}
\label{28} &&S_{11}=\int\!\frac{d^4q}{(2\pi)^4}e^{-iq(x-y)}
\left\{\frac{q_0-E^+}{q_0^2-(E_\Delta^+)^2}\gamma^0\bar\Lambda_++
\frac{q_0+E^-}{q_0^2-(E_\Delta^-)^2}\gamma^0\bar\Lambda_-\right\}
P^{(c)}_{12}+\nonumber\\
&&~~~~~~~~~~~~~~~~~+\int\!\frac{d^4q}{(2\pi)^4}e^{-iq(x-y)}
\left\{\frac{\gamma^0\bar\Lambda_+}{q_0+\breve E^+}+
\frac{\gamma^0\bar\Lambda_-}{q_0-\breve E^-}\right\}P^{(c)}_{3},\\
&&S_{22}= \int\!\frac{d^4q}{(2\pi)^4}e^{-iq(x-y)}
\left\{\frac{q_0-E^-}{q_0^2-(E_\Delta^-)^2}\gamma^0\bar\Lambda_++
\frac{q_0+E^+}{q_0^2-(E_\Delta^+)^2}\gamma^0\bar\Lambda_-\right\}
P^{(c)}_{12}+\nonumber\\
&&~~~~~~~~~~~~~~~~~+\int\!\frac{d^4q}{(2\pi)^4}e^{-iq(x-y)}
\left\{\frac{\gamma^0\bar\Lambda_+}{q_0+\breve E^-}+
\frac{\gamma^0\bar\Lambda_-}{q_0-\breve E^+}\right\}P^{(c)}_{3},
\label{29}\\
&&S_{21}=-i\Delta^*\tau_2\lambda_2
\int\!\frac{d^4q}{(2\pi)^4}e^{-iq(x-y)}
\left\{\frac{\gamma^5\bar\Lambda_+}{q_0^2-
(E_\Delta^+)^2}+
\frac{\gamma^5\bar\Lambda_-}{q_0^2-(E_\Delta^-)^2}\right\},
\label{30}\\
&&S_{12}=-i\Delta\tau_2\lambda_2
\int\!\frac{d^4q}{(2\pi)^4}e^{-iq(x-y)}
\left\{\frac{\gamma^5\bar\Lambda_+}{q_0^2-
(E_\Delta^-)^2}+
\frac{\gamma^5\bar\Lambda_-}{q_0^2-
(E_\Delta^+)^2}\right\},
\label{31}
\end{eqnarray}
where $\bar\Lambda_\pm=\frac
12(1\pm\frac{\gamma^0(\vec\gamma\vec q-m)}{E})$, and
$P^{(c)}_{12}={\rm diag}(1,1,0)$, $P^{(c)}_{3}={\rm
diag}(0,0,1)$ are the projectors on the red/green and blue directions
in the color space, correspondingly.

The poles of the matrix elements (\ref{28})-(\ref{31}) of
the quark propagator give the dispersion laws, i.e. the momentum
dependence of energy, for quarks in a medium. Thus, we have
$E_\Delta^-$ for the energy of red/green quarks and $E_\Delta^+$ for
the energy of red/green antiquarks. Moreover, the
energy of blue quarks (antiquarks) is equal to $\breve E^-$ ($\breve
E^+$). It is clear from Figs 1,2 that in the 2SC phase, i.e. at
$\mu>\mu_c$, we have $(\mu+\mu_8)>m$, $(\mu-2\mu_8)>m$. Then, in this
phase the quantity $E$ can reach both the value $(\mu+\mu_8)$ (the
Fermi energy for red/green quarks) and the value $(\mu-2\mu_8)$ (the
Fermi energy for blue quarks). In this case, in order to create a
red/green quark in the 2SC phase, a minimal amount of energy  (the
gap) equal to $|\Delta|$ at the Fermi level ($E=\mu+\mu_8$) is
required. Similarly, there is no energy cost to create a blue quark
at its Fermi level $E=\mu-2\mu_8$, i.e.  blue quarks are gapless  in
the 2SC phase. In the normal phase of the model, i.e. at $\mu<\mu_c$,
the minimal energy $(m-\mu)$ is needed for the creation of a quark of
any color. Note, both in the 2SC and normal phases of the model the
minimal energy required for a quark creation differs from the minimal
energy required for the creation of an antiquark of the same color.
This fact reflects the breaking of charge conjugation symmetry in
the presence of chemical potentials.

Without loss of generality, we will assume throughout the paper that
$\Delta$ is a real nonnegative quantity. Given the explicit
expression for the quark propagator $S_0$, in the next sections we
will calculate two-point (unnormalized) correlators of meson and
diquark fluctuations over the ground state in the one-loop
(mean-field) approximation and find their masses.

\begin{figure}
%----figure 1 (in the left minipage)
\begin{minipage}[t]{7cm}
 \epsfxsize=80mm   %or \epsfysize= HEIGHT cm
 \centerline{\epsfbox{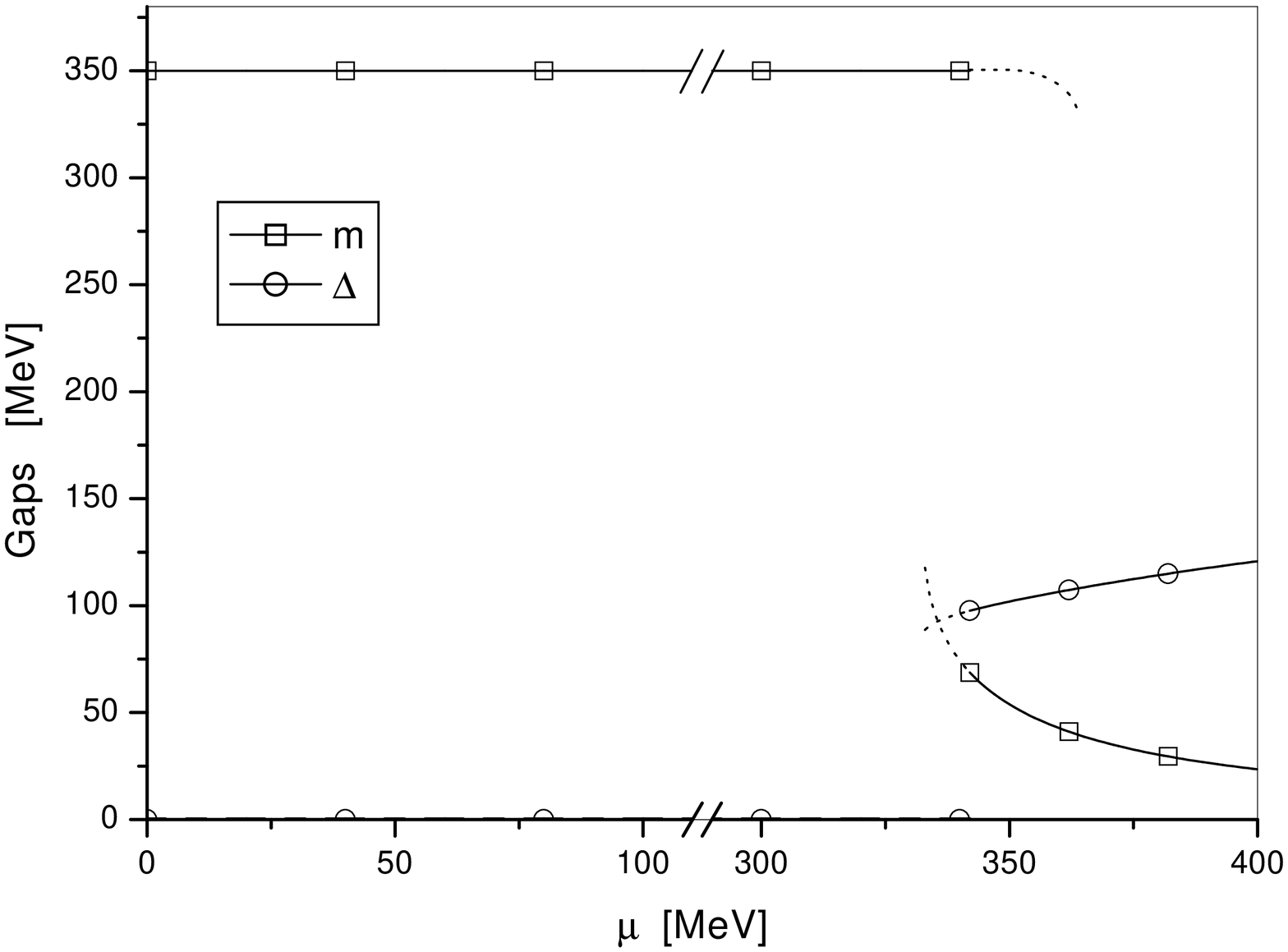}}
 \caption{The constituent quark mass $m$ vs $\mu$ and
color gap $\Delta$ vs $\mu$ under local color neutrality constraint.}
 \label{fig:2}
\end{minipage} \hspace{1cm}
%----figure 2 (in the right minipage)
\begin{minipage}[t]{7cm}
 \epsfxsize= 80mm   %or \epsfysize= HEIGHT cm
 \centerline{\epsfbox{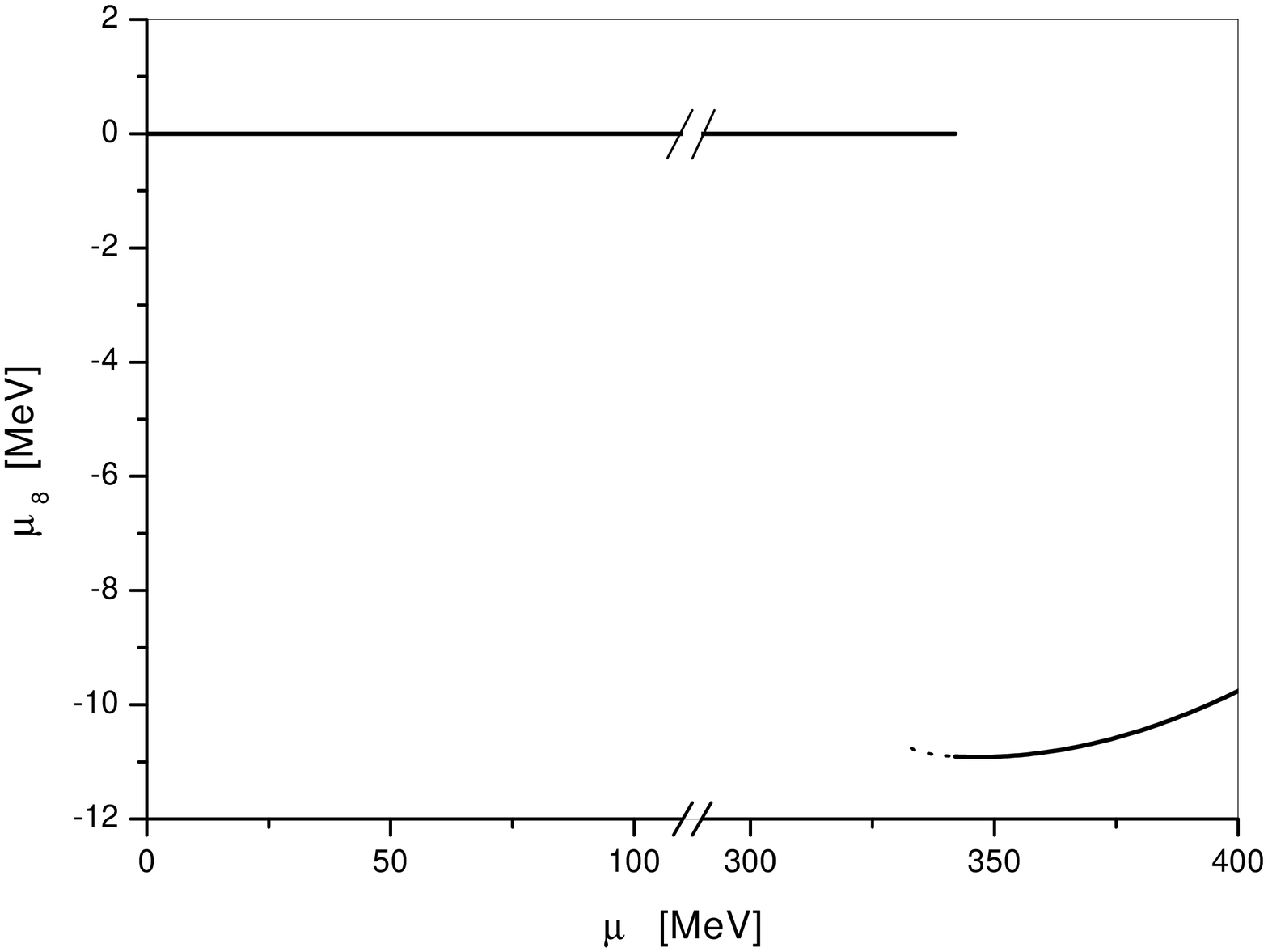}}
\caption{The behaviour of $\mu_8$ vs $\mu$ in the model (\ref{3})
when local color neutrality is imposed.}
\label{fig:3}
\end{minipage}
\end{figure}

\section{Masses of the $\pi$- and $\sigma$-mesons}
\label{IV}

It turns out that the ${\cal S}^{(2)}_{\rm mixed}$ -part (\ref{2.24})
of the effective action is composed from $\sigma (x)$,
$\Delta^s_{2}(x)$ and $\Delta^{s*}_{2}(x)$ fields only. So it
provides us with nondiagonal matrix elements $\Gamma_{\sigma X}$
($X=\Delta^{s*}_2, \Delta^s_2$) of the inverse propagator matrix of
$\sigma$, $\Delta^{s*}_2$, and $\Delta^s_2$. Moreover, each term in
${\cal S}^{(2)}_{\rm mixed}$ is proportional to $\Delta$ or
$\Delta^*$ as well as to the constituent quark mass $m$ (see
Appendix). Hence, in the color symmetric phase ($\Delta =0$) there is
no mixing between $\sigma$-meson and diquark $\Delta^s_{2}$ at all.
The parameter $m$ is small (or even equals zero if $m_0=0$) in the
2SC phase, so we ignore for simplicity the $\sigma$-$\Delta^s_{2}$
mixing effect in this phase, too. As a result, in order to get the
masses of mesons we use only the effective action (\ref{2.22}) which
has the form ${\cal S}^{(2)}_{\rm mesons}={\cal S}^{(2)}_{\rm
\sigma\sigma}+{\cal S}^{(2)}_{\rm \pi\pi}$, where
\begin{eqnarray}
\label{40}
 {\cal S}^{(2)}_{\rm \sigma\sigma} \!\!\!\!&&\!\!=
 -\int d^4x\frac{\sigma^2}{4G}+
\frac i4{\rm Tr}_{scfx}
\left\{S_{11}\sigma S_{11}\sigma +2S_{12}\sigma S_{21}\sigma +
S_{22}\sigma S_{22}\sigma\right\}\equiv \nonumber \\
&&~~~~~~~~~~~~~~~~~~~~~\equiv -\frac 12 \int d^4ud^4v~\sigma
(u)\Gamma(u-v)\sigma (v),\\
{\cal S}^{(2)}_{\rm \pi\pi} \!\!\!\!&&\!\!=
 -\int d^4x\frac{\pi^2_a}{4G}+\frac i4{\rm Tr}_{scfx}
\left\{S_{11}(i\gamma^5\pi_a\tau_a)S_{11}(i\gamma^5\pi_b\tau_b)
+2S_{12}(i\gamma^5\pi_a\tau_a^t)S_{21}(i\gamma^5\pi_b\tau_b) +
\right.\nonumber\\
&&\left.+S_{22}(i\gamma^5\pi_a\tau_a^t)
S_{22}(i\gamma^5\pi_b\tau_b^t)\right\}\equiv -\frac 12 \int
d^4ud^4v~\pi_k (u)\Pi_{kl}(u-v)\pi_l(v).
\label{41}
\end{eqnarray}
In these formulae $\Gamma(x-y)$ is the inverse propagator of
$\sigma$-mesons, and $\Pi_{ab}(x-y)$ is the (diagonal) matrix of the
inverse $\pi$-meson propagator. Evidently, that
\begin{eqnarray}
&& \Gamma(x-y)=-\frac{\delta^2{\cal S}^{(2)}_{\sigma\sigma}}
{\delta\sigma(y)\delta\sigma(x)}, ~~~~~~
\Pi_{ab}(x-y)=-\frac{\delta^2{\cal S}^{(2)}_{\pi\pi}}
{\delta\pi_b(y)\delta\pi_a(x)}.
\label{44}
\end{eqnarray}
Next, using in (\ref{40})-(\ref{41}) the expressions
(\ref{28})-(\ref{31})
for the matrix elements $S_{ij}$, it is possible to obtain with
the help of relations (\ref{44}) the functions $\Gamma(x-y)$,
$\Pi_{ab}(x-y)$, and then their momentum space representations,
$\Gamma(p)$, $\Pi_{ab}(p)$, correspondingly.%
\footnote{One should not be confused by the coordinate or momentum
space representation which is used for the inverse
propagators and other quantities, since this is clear from the
arguments of these functions or the context.}
The zeros of these functions determine the particle and antiparticle
dispersion laws, i.e. the relations between their energy
and three-momenta. In the present paper, we are mainly interested
in the investigation of the modification of meson and diquark masses
in dense and cold color neutral quark matter. In this case, the
particle mass is defined as the value of its energy in the rest
frame, $\vec p=0$ (see, e.g., \cite{eky,ruivo}), where the
calculation of inverse propagators for $\sigma$- and $\pi$-mesons is
significantly simplified. Indeed, in the rest frame it is possible to
get:
\begin{eqnarray}
&&\Pi_{ab}(p_0)=\frac{\delta_{ab}}{2G}
-8\delta_{ab}\int\frac{d^3q}{(2\pi)^3}
\frac{E_\Delta^+E_\Delta^-+E^+E^-+\Delta^2}{E_\Delta^+E_\Delta^-}
\frac{E_\Delta^++E_\Delta^-}{(E_\Delta^++E_\Delta^-)^2-p_0^2}-
\nonumber\\
&&\qquad-16\delta_{ab}\int\frac{d^3q}{(2\pi)^3}
\frac{\theta (E-\breve\mu)E}{4E^2-p_0^2}\equiv
\delta_{ab}\Pi(p_0),
\label{20}
\end{eqnarray}
\begin{eqnarray}
\Gamma(p_0)=\Gamma_0(p_0^2)+\Gamma_1(p_0^2),
\label{22}
\end{eqnarray}
\begin{eqnarray}
&&\Gamma_0(p_0^2)=\frac 1{2G}-
8\int\frac{d^3q}{(2\pi)^3}\,\frac{\vec q^2}{E^2}\,
\frac{E_\Delta^+E_\Delta^-+E^+E^-+\Delta^2}{E_\Delta^+E_\Delta^-}
\,\frac{E_\Delta^++E_\Delta^-}{(E_\Delta^++E_\Delta^-)^2-p_0^2}
-16\int\frac{d^3q}{(2\pi)^3}\,\frac{\vec q^2}{E}\,
\frac{\theta(E-\breve \mu)}{4E^2-p_0^2},\label{220}\\
&&\Gamma_1(p_0^2)=16\Delta^2m^2\int\frac{d^3q}{(2\pi)^3
E^2}\,
\left\{\frac{1}{E_\Delta^+[p_0^2-4(E_\Delta^+)^2]}
+\frac{1}{E_\Delta^-[p_0^2-4(E_\Delta^-)^2]}\right\},
\label{221}
\end{eqnarray}
where the same notations as in (\ref{32}) were used. The zeros of
the functions (\ref{20}), (\ref{22}) give us the masses of $\pi$- and
$\sigma$-mesons, respectively (see Fig. 3). \footnote{In our
numerical investigations of the 2SC phase, presented in Fig. 3,
we have ignored in (\ref{22}) the term $\Gamma_1(p_0^2)$ proportional
to $\Delta^2m^2$, since it is comparable (or even less) in magnitude
with nondiagonal elements $\Gamma_{\sigma X}(p_0)$
($X=\Delta^{s*}_2, \Delta^s_2$) of the full inverse propagator matrix
of $\sigma$, $\Delta^{s*}_2$, and $\Delta^s_2$ (see Appendix), which
are not taken into account in the above consideration.} We see that
in the 2SC phase the masses of the sigma- and pi-meson are about 300
MeV. Moreover, the pion is a stable particle in this phase (only
electroweak decay channels are allowed). This conclusion is supported
by the following arguments. It is clear that the first and the second
integrals in (\ref{20}) are analytical functions in the whole complex
$p_0^2$ plane, except the cuts $E_{\rm min}^2<p_0^2<\infty$ and
$(2\breve\mu)^2<p_0^2<\infty$, respectively (here $E_{\rm min}=
\sqrt{(\bar\mu-m)^2+|\Delta|^2}+\sqrt{(\bar\mu+m)^2+|\Delta|^2}$ is
the minimum of the expression $E_\Delta^-+E_\Delta^+$, which is taken
at the point $|\vec p|=0$). Evidently, $E_{\rm min}$ corresponds to
the threshold for the pion decay into a red-green quark-antiquark
pair, whereas $2\breve\mu$ corresponds to the threshold for the pion
decay into a blue quark-antiquark pair. It is easily seen from Fig. 3
that in the 2SC phase the pion mass is less than the values of these
two thresholds. Because there are no other singularities in
(\ref{20}) corresponding to different channels of the pion decay, we
can conclude that in the 2SC phase the pion is a stable particle.

The neglection of the
mixing between the $\sigma$-meson and $\Delta^s_2$-scalar diquark
also results in a stable $\sigma$-meson. However, if the mixing is
taken into account, then in the 2SC phase the $\sigma$-meson is a
resonance, decaying into a pair of quarks, whose width is a rather
small quantity, i.e. about 30 MeV (see Appendix).
\begin{figure}
  \centering
  \includegraphics[width=14cm]{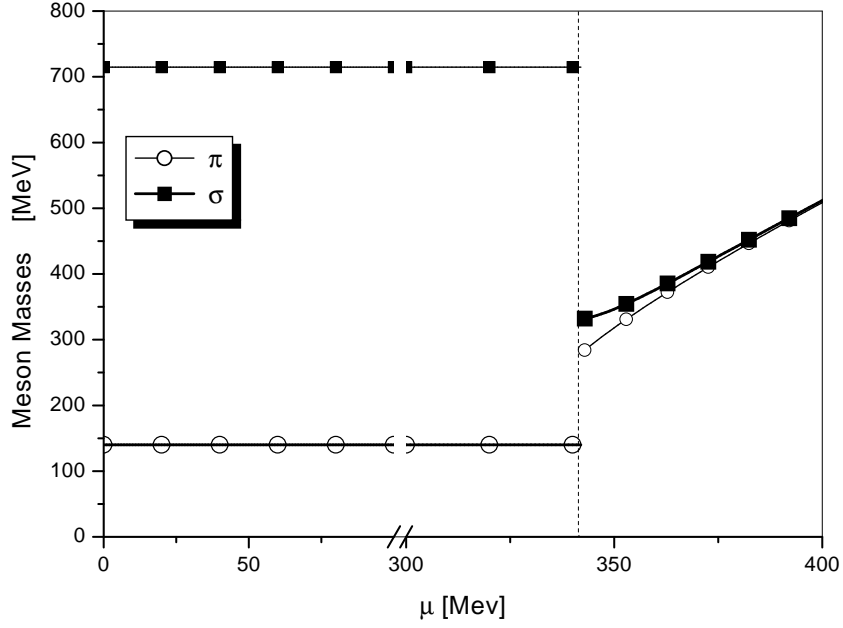}
  \caption{The masses of the $\sigma$-meson  and pion
   as functions of $\mu$, when mixing of $\sigma$ and
  $\Delta^s_2$ is neglected.}\label{plot:mesonmasses}
\end{figure}

\section{Masses of scalar diquarks}
\label{V}

As in the previous section, we will ignor for simplicity the
term ${\cal S}^{(2)}_{\rm mixed}$ (\ref{2.24}), which mixes $\sigma$
and $\Delta^{s*}_2, \Delta^s_2$ diquarks, in the effective
action (\ref{2.21}). In this case, in order to obtain the masses of
diquarks, we need to analyze the term ${\cal S}^{(2)}_{\rm diquarks}$
(\ref{2.23}) only. It can be easily presented in the following form
\begin{eqnarray}
\label{3.36}
{\cal S}^{(2)}_{\rm diquarks}=\sum_{A=2,5,7}\Big\{{\cal S}^{(2)}_{\rm
s~AA}+{\cal S}^{(2)}_{\rm p~AA}\Big\},
\end{eqnarray}
where labels $s,p$ denote the contributions from scalar- and
pseudo-scalar diquark fields, correspondingly, and
\begin{eqnarray}
\label{3.37}
{\cal S}^{(2)}_{\rm s~AA}\!\!\!\!&&\!\!=
 -\int d^4x\frac{\Delta^s_{A}\Delta^{s*}_{A}}{4H_s}+
\frac i2{\rm Tr}_{scfx}
\left\{S_{11}i\Delta^s_{A}\gamma^5\tau_2\lambda_{A}S_{22}
i\Delta^{s*}_{A}\gamma^5\tau_2\lambda_{A}
\right\},\\
\label{3.370}
{\cal S}^{(2)}_{\rm p~AA}\!\!\!\!&&\!\!=
 -\int d^4x\frac{\Delta^p_{A}\Delta^{p*}_{A}}{4H_p}
+\frac i2{\rm Tr}_{scfx}
\left\{S_{11}\Delta^p_{A}\tau_2\lambda_{A}S_{22}
\Delta^{p*}_{A}\tau_2\lambda_{A}\right\},
\end{eqnarray}
for fixed $A=5,7$, and
\begin{eqnarray}
 {\cal S}^{(2)}_{\rm s~22} \!\!\!\!&&\!\!=
 -\int d^4x\frac{\Delta^s_{2}\Delta^{s*}_{2}}{4H_s}+
\frac i4{\rm Tr}_{scfx}
\left\{S_{12}i\Delta^{s*}_{2}\gamma^5\tau_2
\lambda_{2}S_{12}i\Delta^{s*}_{2}\gamma^5\tau_2
\lambda_{2}+\right.\nonumber\\
&&\left.~~~~~~~~+2S_{11}i\Delta^s_{2}\gamma^5\tau_2
\lambda_{2}S_{22}i\Delta^{s*}_{2}\gamma^5\tau_2
\lambda_{2}+S_{21}i\Delta^s_{2}\gamma^5\tau_2
\lambda_{2}S_{21}i\Delta^s_{2}\gamma^5\tau_2
\lambda_{2}\right\},
\label{3.38}\\
 {\cal S}^{(2)}_{\rm p~22} \!\!\!\!&&\!\!=
 -\int d^4x\frac{\Delta^p_{2}\Delta^{p*}_{2}}{4H_p}
+\frac i4{\rm Tr}_{scfx}
\left\{S_{12}\Delta^{p*}_{2}\tau_2
\lambda_{2}S_{12}\Delta^{p*}_{2}\tau_2
\lambda_{2}+\right.\nonumber\\
&&\left.~~~~~~~~~~~~~~+2S_{11}\Delta^p_{2}\tau_2
\lambda_{2}S_{22}\Delta^{p*}_{2}\tau_2
\lambda_{2}+S_{21}\Delta^p_{2}\tau_2
\lambda_{2}S_{21}\Delta^p_{2}\tau_2
\lambda_{2}\right\}.
\label{3.380}
\end{eqnarray}
It follows from (\ref{3.36})-(\ref{3.380}) that there is no mixing
between scalar- and pseudo-scalar diquarks. Moreover, scalar diquarks
(or pseudo-scalar ones), as such, are not mixed to one another.
Starting from the above formulae, we will find the inverse
propagators of diquarks which migth be introduced by the following
way ($A=2,5,7$):
\begin{eqnarray}
&&{\cal S}^{(2)}_{s~AA}=
-\frac 12 \sum_{X,Y}\int d^4ud^4v~X(u)
\Gamma^{As}_{XY}(u-v)Y(v),\label{3.430}\\
&&{\cal S}^{(2)}_{p~AA}=
-\frac 12 \sum_{P,Q}\int d^4ud^4v~P(u)
\Gamma^{Ap}_{PQ}(u-v)Q(v),
\label{3.43}
\end{eqnarray}
where (for each fixed value of $A$)
$X(x),Y(x)=\Delta^s_A(x),\Delta^{s*}_A(x)$, $P(x),Q(x)=\Delta^p_A(x),
\Delta^{p*}_A(x)$ and $\Gamma^{As}_{XY}(z)$ or $\Gamma^{Ap}_{PQ}(z)$
are matrix elements of the $2\times 2$ inverse propagator matrix for
$\Delta^s_A(x),\Delta^{s*}_A(x)$- or $\Delta^p_A(x),
\Delta^{p*}_A(x)$-fields, respectively. Given diquark propagators, it
is possible then to obtain the masses of diquarks.

\subsection{Scalar diquarks in the 2SC phase ($\Delta\ne 0$,
$\mu_8\ne 0$)}
\label{VA}

In the present section, using (\ref{3.430}) for different $A=2,5,7$,
we will study step-by-step the masses of scalar diquark excitations
in the 2SC phase of the model (\ref{3}), when the color neutrality
condition is taken into account. Let us begin with the $\Delta^s_5$
-$\Delta^{s*}_5$ diquark sector. It follows from (\ref{3.430}) at
$A=5$ that
\begin{eqnarray}
&& \Gamma^{5s}_{XY}(x-y)=-\frac{\delta^2{\cal S}^{(2)}_{s~55}}
{\delta Y(y)\delta X(x)}
\label{3.44}
\end{eqnarray}
(recall, $X,Y=\Delta^s_5(x),\Delta^{s*}_5(x)$ in the case under
consideration). Note also that $\Gamma^{5s} (z)$ is a symmetric
matrix, i.e. $\Gamma^{5s}_{XY}(z)=\Gamma^{5s}_{YX}(-z)$. It is clear
from (\ref{3.430}) and (\ref{3.44}) that
$\Gamma^{5s}_{\Delta^s_5\Delta^s_5}(z)=\Gamma^{5s}_{\Delta^{s*}_5
\Delta^{s*}_5}(z)=0$, and this matrix has nonzero elements of the
form (in the momentum space  representation):
\begin{eqnarray}
&&\Gamma^{5s}_{\Delta^{s*}_5\Delta^s_5}(p)=\frac{1}{4H_s}-
i{\rm Tr}_{sc}\int\frac{d^4q}{(2\pi)^4}
\left\{S_{11}(q+p)i\gamma^5\lambda_5
S_{22}(q)i\gamma^5\lambda_5
\right\},
\label{3.49}\\
&&\Gamma^{5s}_{\Delta^s_5\Delta^{s*}_5}(p)=\frac{1}{4H_s}-
i{\rm Tr}_{sc}\int\frac{d^4q}{(2\pi)^4}
\left\{S_{22}(q+p)i\gamma^5\lambda_5
S_{11}(q)i\gamma^5\lambda_5
\right\},
\label{3.50}
\end{eqnarray}
where the Fourier-transformed expressions $S_{11}(q)$,
$S_{22}(q)$ can be easily determined from (\ref{28}),
(\ref{29}). It follows from (\ref{3.49})-(\ref{3.50}) that
$\Gamma^{5s}_{\Delta^{s*}_5\Delta^s_5}(-p)=$
$\Gamma^{5s}_{\Delta^s_5\Delta^{s*}_5}(p)$. Since we are
interested in diquark masses, it is necessary to use the rest frame
in (\ref{3.49})-(\ref{3.50}), i.e. $p=(p_0,0,0,0)$ (see also
\cite{eky,ruivo}). In this case the calculation of matrix elements
(\ref{3.49})-(\ref{3.50}) is greatly simplified, and mass excitations
are connected with the zeros of the quantity ${\rm
det}~\Gamma^{5s} (p_0)$ in the $p_0^2$-plane.%
\footnote{Recently,  the Bethe-Salpeter
equation approach has been used to obtain diquark masses in the 2SC
phase of cold dense QCD at asymptotically large values of the
chemical potential \cite{mirsh}. There, the mass of the diquark was
defined as the energy of a bound state of two virtual quarks in
the center of mass frame, \textit{i.~e.\/}  as in our approach, in
the rest frame for the whole diquark.} So, we have
\begin{eqnarray}
&&\Gamma^{5s}_{\Delta^{s*}_5\Delta^s_5}(p_0)=\frac{1}{4H_s
}-4i\int\frac{d^4q}{(2\pi)^4}
\left\{\frac{q_0+E^+}{(p_0+q_0+\breve E^+)(q_0^2-(E_\Delta^+)^2)}+
\frac{q_0-E^-}{(p_0+q_0-\breve
E^-)(q_0^2-(E_\Delta^-)^2)}\right\}.
\label{3.53}
\end{eqnarray}
The $p_0,q_0$-dependency in the integrand of (\ref{3.53}) is
presented in an evident form. Other quantities in (\ref{3.53}), such
as $E^\pm$ etc, depend on $|\vec q|$ only. The expression
(\ref{3.53}) is  valid both for $\Delta =0$ and $\Delta\ne 0$. For
the case $\Delta\ne 0$, i.e. in the color superconducting phase, it
is possible to use the gap equation (\ref{33}) in order to eliminate
the coupling constant $H_s$ from this formula. In this way we find:
\begin{eqnarray}
&&\Gamma^{5s}_{\Delta^{s*}_5\Delta^s_5}(p_0)=4i(p_0-3\mu_8
)\int\frac{d^4q}{(2\pi)^4}
\left\{\frac{1}{(p_0+q_0+\breve
E^+)(q_0^2-(E_\Delta^+)^2)}+\frac{1}{(p_0+q_0-\breve E^-)
(q_0^2-(E_\Delta^-)^2)}\right\}\equiv\nonumber\\
&&~~~~~~~\equiv 2(p_0-3\mu_8)H(p_0),~~~~~~~~~~~~~
\Gamma^{5s}_{\Delta^s_5\Delta^{s*}_5}(p_0)
=-2(p_0+3\mu_8)H(-p_0).
\label{3.55}
\end{eqnarray}
The last equation in (\ref{3.55}) is due to the relation
$\Gamma^{5s}_{\Delta^{s*}_5\Delta^s_5}(-p_0)=$
$\Gamma^{5s}_{\Delta^s_5\Delta^{s*}_5}(p_0)$. Recall, in
(\ref{3.55}) $q_0+p_0$ and $q_0$ are shorthands for
$(p_0+q_0)+i\varepsilon\cdot {\rm sign}(p_0+q_0)$ and
$q_0+i\varepsilon\cdot {\rm sign}(q_0)$, where $\varepsilon\to 0_+$
(see also comment after formula (\ref{34})). Performing in
(\ref{3.55}) the $q_0$-integration, one obtains
\begin{eqnarray}
H(p_0)=\int\frac{d^3q}{(2\pi)^3}
\left\{\frac{\theta(\breve E^+)}{(p_0+\breve
E^++E_\Delta^+)E_\Delta^+}\!\!\right.&+&\!\!
\frac{\theta(-\breve E^+)}{(p_0+\breve E^+-E_\Delta^+)E_\Delta^+}+
\frac{\theta(\breve E^-)}{(p_0-\breve E^--E_\Delta^-)E_\Delta^-}+
\nonumber\\
&&\left.+\frac{\theta(-\breve E^-)}{(p_0-\breve
E^-+E_\Delta^-)E_\Delta^-}\right\}.
\label{3.56}
\end{eqnarray}
Now, with the help of (\ref{3.55}) one easily gets the expression for
the determinant of the inverse propagator matrix
$\Gamma^{5s} (p_0)$ in the rest frame:
\begin{eqnarray}
{\rm det}\Gamma^{5s} (p_0)=
\Gamma^{5s}_{\Delta^{s*}_5\Delta^s_5}(p_0)
\Gamma^{5s}_{\Delta^s_5\Delta^{s*}_5}(p_0)
\equiv -4(p_0^2-9\mu_8^2)H(p_0)H(-p_0).
\label{3.57}
\end{eqnarray}
 The diquark squared mass spectrum in the $\Delta^s_5$-sector of the
 theory is defined by zeros of the det$\Gamma^{5s} (p_0)$
 in the $p_0^2$-plane. Evidently, the point $p_0^2=9\mu_8^2$ is the
 solution of the equation det$\Gamma^{5s} (p_0)=0$. Let us
 suppose that  some nonzero point $p_0=-M^s_{D}$ is the zero of the
 function  $H(p_0)$, i.e. $H(-M^s_{D})=0$. Then, at $p_0=M^s_{D}$ the
 determinant  (\ref{3.57}) is also equal to zero, so the point
 $p_0^2=(M^s_{D})^2$ is  another zero of ${\rm det} \overline
 {\Gamma^{5s}}(p_0)$ in the  $p_0^2$-plane, and the second bosonic
 excitation  of this sector has  nonzero mass $M^s_{D}$. It follows
 from (\ref{3.56}) and (\ref{35}) that
 $H(p_0)$ is proportional to $\vev{Q_8}$ at the point $p_0=3\mu_8$.
 Namely, $H(3\mu_8)\equiv -\vev{Q_8}\big /(4\Delta^2)$. Since
 $\vev{Q_8}$ is zero due to the constraint (\ref{35}), we may
 conclude that $M^s_{D}=3|\mu_8|$. Hence, in the $\Delta^s_5$-sector
 of the  model there are two real bosonic excitations with equal
 masses  $M^s_{D}\equiv 3|\mu_8|$.

The similar is true for the $\Delta^s_7$-sector of the model, so in
the whole $\Delta^s_5, \Delta^s_7$-sector of the NJL model that is
under the color neutrality constraint there are four massive  scalar
excitations with equal masses $M^s_{D}\equiv 3|\mu_8|$. These
particles form two real antidoublets of the $\rm SU_c(2)$-group, or
one complex antidoublet.

Consider now the diquark excitations of the 2SC ground state in the
$\Delta^s_2$-$\Delta^{s*}_2$ sector of the model. In this case, the
matrix $\Gamma^{2s} (p_0)$ (the momentum representation
for the inverse propagator matrix at $\vec p=0$, i.e. in the rest
frame) has the following structure:
\begin{eqnarray}
\Gamma^{2s}_{\Delta^s_2\Delta^s_2}(p_0)&=&\Gamma^{2s}_{\Delta^{s*}_2
\Delta^{s*}_2}(p_0)=
4\Delta^2I_0(p_0^2),~~~~~~~\Gamma^{2s}_{\Delta^s_2\Delta^{s*}_2}(p_0)
=\Gamma^{2s}_{\Delta^{s*}_2\Delta^s_2}(-p_0)=\nonumber\\
&=&(4\Delta^2-2p_0^2)I_0(p_0^2)+4p_0I_1(p_0^2),
\label{24}
\end{eqnarray}
where
\begin{eqnarray}
&&I_0(p_0^2)=\int\frac{d^3q}{(2\pi)^3}
\frac{1}{E_\Delta^+[4(E_\Delta^+)^2-p_0^2]}+\int
\frac{d^3q}{(2\pi)^3}
\frac{1}{E_\Delta^-[4(E_\Delta^-)^2-p_0^2]},\label{25}\\
&&I_1(p_0^2)=\int\frac{d^3q}{(2\pi)^3}
\frac{E^+}{E_\Delta^+[4(E_\Delta^+)^2-p_0^2]}-
\int\frac{d^3q}{(2\pi)^3}
\frac{E^-}{E_\Delta^-[4(E_\Delta^-)^2-p_0^2]}.\label{26}
\end{eqnarray}
The mass spectrum is defined by the equation
\begin{equation}
  \label{032}
{\rm det}\Gamma^{2s} (p_0)\equiv
4p_0^2\big\{(p_0^2-4\Delta^2 )I_0^2(p_0^2)-4I_1^2(p_0^2)\big\}=0.
\end{equation}
In the $p_0^2$-plane this equation has an evident zero, corresponding
to a Nambu--Goldstone boson,  $p_0^2=0$. Detailed investigation,
similar to that from \cite{eky}, shows that on the second Riemann
sheet of $p_0^2$ there is another zero of (\ref{032}) which
corresponds to a heavy resonance. Its mass $M$ and width $\Gamma$ are
depicted in the Fig. 4.

As a result, we conclude that in the 2SC phase there are four light
real scalar diquark excitations with mass $3|\mu_8|$, and one
Nambu--Goldstone boson, which appears due to a spontaneous breaking
of the $\rm SU_c(2)\times U_{\lambda_8}(1)$ color symmetry down to
$\rm SU_c(2)$. Moreover, a heavy scalar diquark resonance, which is
an $\rm SU_c(2)$-singlet, is also presented in the mass spectrum of
the model at  $\mu>\mu_c =342$ MeV (see Fig. 4).

\begin{figure}
  \centering
  \includegraphics[width=14cm]{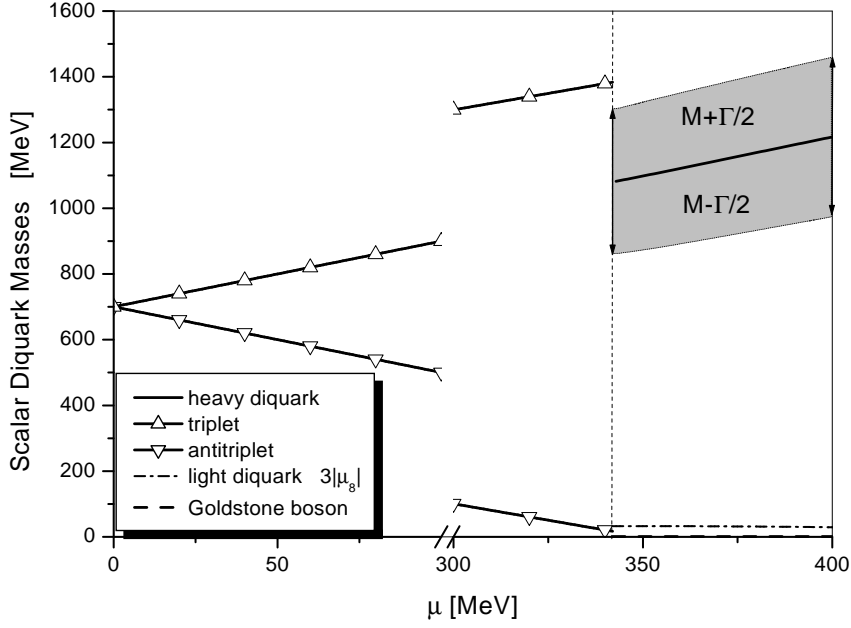}
  \caption{The masses of diquarks. At $\mu<\mu_c=342$ MeV, six
  diquark states are splitted into a (color)triplet of heavy states
  with mass $M_{\Delta^*}$ and an anti-triplet of light states with
  mass $M_\Delta$ (see (\ref{300})). In the 2SC phase
($\mu>\mu_c$), one observes the Numbu--Goldstone boson, four light
diquarks with the mass $3|\mu_8|$, and a heavy
singlet state with the mass $M$ (solid line). The shaded
rectangular displays the width $\Gamma$ of the heavy singlet
resonance; its upper border is half-width higher than the mass and
the bottom border is half-width lower.}\label{plot:diquarkmasses}
\end{figure}

\subsection{Scalar diquarks in the normal phase ($\Delta =0$,
$\mu_8=0$)}\label{VB}

In the $\rm SU_c(3)$-symmetric phase ($\mu<\mu_c$) the diquark gap
$\Delta$ is zero and the three complex diquark fields $\Delta^s_A(x)$
($A=2,5,7$) are not mixed with other fields in the second order
effective action of the model. Moreover, at $\Delta =0$, as it is
easily seen from the color neutrality constraint (\ref{35}), we have
$\mu_8=0$. So, in order to study the diquark masses, it is enough to
consider, e.g., the $\Delta^s_5$-diquark sector. In this phase the
determinant of the inverse propagator matrix $\Gamma^{5s} (p_0)$
looks like (we use the rest frame, where $\vec p =0$)
\begin{eqnarray}
{\rm det}\Gamma^{5s}(p_0)
=\Gamma^{5s}_{\Delta^{s*}_5\Delta^s_5}(p_0)
\Gamma^{5s}_{\Delta^s_5\Delta^{s*}_5}(p_0)=
\Gamma^{5s}_{\Delta^{s*}_5\Delta^s_5}(p_0)
\Gamma^{5s}_{\Delta^{s*}_5\Delta^s_5}(-p_0),
\label{det2}
\end{eqnarray}
where $\Gamma^{5s}_{\Delta^{s*}_5\Delta_5}(p_0)$ is
presented in (\ref{3.53}). (Note, the last equality in formula
(\ref{3.57}) for ${\rm det}\Gamma^{5s}(p_0)$ in the 2SC
phase is based on the usage of a non trivial solution $\Delta\ne 0$
of the gap equation (\ref{33}). So, in the normal $\rm SU_c(3)
$-symmetric phase, where $\Delta =0$, it is not valid.) Taking into
account the relation $m>\mu$ that is realized in the normal phase
only (see Fig. 1), we see that $E\equiv \sqrt{\vec q^2+m^2}>\mu$ and,
as a consequence, $E^\pm>0$ are fulfilled in this phase. So, one can
easily integrate in (\ref{3.53}) over $q_0$ and obtain the
following expression that is suitable only for $\rm SU_c(3)
$-symmetric phase:
\begin{eqnarray}
\Gamma^{5s}_{\Delta^{s*}_5\Delta^s_5}(p_0)= \frac
1{4H_s}-16
\int\frac{d^3q}{(2\pi)^3}\frac{E}{4E^2-(p_0+2\mu)^2}\equiv
\frac 1{4H_s}-F_s(\epsilon),
\label{gamma}
\end{eqnarray}
where $\epsilon=(p_0+2\mu)^2$.
Clearly, the diquark mass spectrum is defined by the equation ${\rm
det}\Gamma^{5s}(p_0)=0$, or by zeros of (\ref{gamma}),
where the function $F_s(\epsilon)$ is analytical in
the whole complex $\epsilon$-plane, except for the cut
$4m^2<\epsilon$ along the real axis. (In general, $F_s(\epsilon)$ is
defined on a complex Riemann surface which is to
be described by several sheets. However, a direct numerical
computation based on eq.\ (\ref{gamma}) gives its values on the first
sheet only (we use the parameter set (\ref{350})).
To find a value on the rest of the Riemann surface, a special
procedure of continuation is needed.) The numerical analysis of
(\ref{gamma}) on the first Riemann sheet shows  that the equation
$\Gamma^{5s}_{\Delta^{s*}_5\Delta_5}(p_0)=0$ has
a root ($\epsilon_0$) on the real axis ($0<\epsilon_0<4m^2$),
providing us with the following massive diquark modes which are the
solutions of the eq. (\ref{det2})
\begin{equation}
  \label{300}
(M^s_{\Delta})^2=(1.998 m-2\mu)^2,~~~~~(M^s_{\Delta^{*}})^2=
(1.998 m+2\mu)^2.
\end{equation}
We relate   $M^s_{\Delta}$
in (\ref{300}) to the mass of the diquark with the baryon
number $B=2/3$ and $M^s_{\Delta^{*}}$ to the mass of
the antidiquark with $B=-2/3$. (Qualitatively, a similar
behavior of diquark and antidiquark masses vs. $\mu$ was
obtained in \cite{ratti} in the NJL model with two-colored quarks.)
The difference between diquark and antidiquark masses in (\ref{300})
is explained by the absence of a charge conjugation symmetry in the
presence of a chemical potential.

Finally,  due to the underling color $\rm SU(3)_c$ symmetry, the
previous statement is valid also for $\Delta^*_5, \Delta_5$ and
$\Delta^*_7, \Delta_7$. As a result, we have a color antitriplet of
diquarks  with the mass $M^s_{\Delta}$ (\ref{300})  as well as a
color triplet of antidiquarks with the mass $M^s_{\Delta^{*}}$. The
results of numerical computations are presented in Fig.~4 for
$\mu<\mu_c =342$~MeV.

Recall that in our analysis, we have used the constraint $H_s=3G/4$,
thereby fixing the constant $H_s$ through $G$. It is useful, however,
to discuss now the influence of  $H_s$  on diquark masses. Indeed, it
is clear from (\ref{gamma}) that the root $\epsilon_0$  lies inside
the interval $0<\epsilon_0<4m^2$ only if $H_s^*<H_s<H_s^{**}$, where
$H_s^*$ and $H_s^{**}$ are defined by
\begin{eqnarray}
H_s^* &\equiv&\frac {1}{4F_s(4m^2)}= \frac {\pi^2}{4\left
[\Lambda\sqrt{m^2+\Lambda^2}+m^2\ln((
\Lambda+\sqrt{m^2+\Lambda^2})/m)\right ]},\nonumber\\
H_s^{**}&\equiv&\frac {1}{4F_s(0)}=\frac {\pi^2}{4\left
[\Lambda\sqrt{m^2+\Lambda^2}-m^2\ln((
\Lambda+\sqrt{m^2+\Lambda^2})/m)\right ]}=\frac{3mG}{2(m-m_0)}.
\label{gamma4}
\end{eqnarray}
In this case, there are stable diquarks and antidiquarks in the color
symmetric phase. The behavior of their masses qualitatively resembles
that given by eqs.~(\ref{300}). For a rather weak interaction in the
diquark channel ($H_s<H_s^*$),  $\epsilon_0$ runs onto the second
Riemann sheet, and unstable diquark modes (resonances) appear. Unlike
this, a sufficiently strong interaction in the diquark channel
($H_s>H_s^{**}$) pushes $\epsilon_0$ towards the negative
semi-axis, \textit{i.~e.\/}\ $(p_0+2\mu)^2<0$. The latter indicates a
tachyon singularity in the diquark propagator, evidencing that
the $\rm SU_c(3)$-color symmetric ground state is not stable. Indeed,
at a very large  $H_s$, as it has been shown in \cite{ek} at $H_p=0$,
the color symmetry is spontaneously broken even at a vanishing
chemical potential. We guess that this result remains true at rather
small values of $H_p$ ($H_p<H_s$) as well, justifying the above
mentioned tachyon singularity of the diquark propagator.

\section{Masses of pseudo-scalar diquarks}
\label{VI}

It is clear from (\ref{3.36}) that in the framework of the NJL model
(\ref{3}) pseudo-scalar diquarks are not mixing with each other
as well as with meson- and scalar diquark fields. So, to get their
masses, we will start from the general expression (\ref{3.43}) for
the inverse propagator matrices of pseudo-scalar diquarks.

\subsection{Pseudo-scalar diquarks in the 2SC phase ($\Delta\ne 0$,
$\mu_8\ne 0$)}\label{VIA}

In the 2SC phase there is an $\rm SU(2)_c$ symmetry between
$\Delta^p_5$-$\Delta^{p*}_5$ and $\Delta^p_7$-$\Delta^{p*}_7$
sectors, so it is enough to study an inverse propagator, e.g., in one
of these sectors. For $A=5$ it follows from (\ref{3.43}) that
\begin{eqnarray}
&& \Gamma^{5p}_{PQ}(x-y)=-\frac{\delta^2{\cal S}^{(2)}_{p~55}}
{\delta Q(y)\delta P(x)},
\label{3.59}
\end{eqnarray}
where $P(x),Q(x)$ = $\Delta^p_5(x),\Delta^{p*}_5(x)$. Further, using
the evident expression (\ref{3.370}) for ${\cal S}^{(2)}_{p~55}$,
one can obtain the matrix elements (\ref{3.59}) in the rest frame,
$\vec p=0$, of the momentum space representation
\begin{eqnarray}
&&\Gamma^{5p}_{\Delta^{p*}_5\Delta^p_5}(p_0)=\frac{1}{4H_p}
+4\int\frac{d^3q}{(2\pi)^3}
\left\{\frac{m^2}{E^2}\Big [\frac{(\breve E^--p_0+E^+)\theta(-\breve
E^-)}{(p_0-\breve E^-)^2 -(E_\Delta^+)^2}-
\frac{E^+-E_\Delta^+}{2(p_0-\breve E^-
-E_\Delta^+)E_\Delta^+}-\right.\nonumber\\
&&~~~~~~~~-\frac{(\breve E^++p_0+E^-)\theta(\breve
E^+)}{(p_0+\breve E^+)^2 -(E_\Delta^-)^2}+
\frac{E^-+E_\Delta^-}{2(p_0+\breve E^+ -E_\Delta^-)E_\Delta^-}
\Big ]+\frac{\vec q^2}{E^2}\Big [
\frac{(E^+-\breve E^+-p_0)\theta(\breve
E^+)}{(p_0+\breve E^+)^2 -(E_\Delta^+)^2}-\nonumber\\
&&~~~~~~~~\left.-\frac{E^+-E_\Delta^+}{2(p_0+\breve E^+
-E_\Delta^+)E_\Delta^+}
+\frac{(\breve E^--p_0-E^-)\theta(-\breve
E^-)}{(p_0-\breve E^-)^2 -(E_\Delta^-)^2}+
\frac{E^-+E_\Delta^-}{2(p_0-\breve E^--E_\Delta^-)E_\Delta^-}\Big ]
\right\},
\label{3.66}
\end{eqnarray}
$\Gamma^{5p}_{\Delta^p_5\Delta^{p*}_5}(p)=
\Gamma^{5p}_{\Delta^{p*}_5\Delta^p_5}(-p)$, and
$\Gamma^{5p}_{\Delta^p_5\Delta^p_5}(p)=$
$\Gamma^{5p}_{\Delta^{p*}_5\Delta^{p*}_5}(p)=0$.
Let $\frac 1{4H_p}=\frac 1{4H_s}+\eta$. Then, using for $\frac
1{4H_s}$ the gap equation (\ref{33}) (recall, in the 2SC phase
$\Delta\ne 0$), we obtain from (\ref{3.66})
(transforming the multiplier before the second square bracket in
(\ref{3.66}) as $\vec q^2/E^2=1-m^2/E^2$):
\begin{eqnarray}
&&\Gamma^{5p}_{\Delta^{p*}_5\Delta^p_5}(p_0)=\eta +
2(p_0-3\mu_8)H(p_0)+m^2\widetilde H(p_0),
\label{3.67}
\end{eqnarray}
where $H(p_0)$ is defined in (\ref{3.56}). Since in the 2SC phase
$m<<\Delta$ (or $m$ even a zero if $m_0=0$), we ignore the last term
in (\ref{3.67}) (due to this reason an explicit form of $\widetilde
H(p_0)$ is not presented here) and obtain in this way the
following expression for the determinant of the matrix
$\Gamma^{5p} (p_0)$:
\begin{eqnarray}
&&{\rm det}\Gamma^{5p} (p_0)=
\Gamma^{5p}_{\Delta^{p*}_5\Delta^p_5}(p_0)
\Gamma^{5p}_{\Delta^p_5\Delta^{p*}_5}(p_0)=
\Gamma^{5p}_{\Delta^{p*}_5\Delta^p_5}(p_0)
\Gamma^{5p}_{\Delta^{p*}_5\Delta^p_5}(-p_0)=\nonumber\\
&&\Big [\eta+2(p_0-3\mu_8)H(p_0)\Big ]
\Big [\eta-2(p_0+3\mu_8)H(-p_0)\Big ]\approx
\Big [\eta+2(p_0-3\mu_8)^2\beta\Big ]
\Big [\eta+2(p_0+3\mu_8)^2\beta\Big ],
\label{3.69}
\end{eqnarray}
where $\beta=dH(p_0)/dp_0|_{p_0=3\mu_8}$. In the last relation in
(\ref{3.69}) we have expanded the function $H(p_0)$ into a
Taylor-series of $p_0$ at the point $p_0=3\mu_8$, and took into
account that $H(3\mu_8)$ equals zero under the color neutrality
constraint (see the end of section \ref{VA}). Solving in this
approximation the equation ${\rm det}\Gamma^{5p} (p_0)=0$, it is
possible to find pseudo-scalar diquark excitations with two different
masses:
\begin{eqnarray}
M^p_{D1}=\Big |~3\mu_8+\sqrt{-\eta/(2\beta)}~\Big |,
~~~~~~~~~~~~~~
M^p_{D2}=\sqrt{-\eta/(2\beta)}-3\mu_8.
\label{3.70}
\end{eqnarray}
Since $\beta <0$ (as is easily seen from (\ref{3.56})), we conclude
from formulae (\ref{3.70}) that at $\eta>0$ both $M^p_{D1}$ and
$M^p_{D2}$ are real (positive) quantities, suggesting that these
pseudo-scalars are stable particles at $H_p\le H_s$. (The case
$\eta<0$, which corresponds to an unstable 2SC ground state, will be
discussed in details below, after (\ref{3.90}).) The similar is true
in the $\Delta^p_7$-$\Delta^{p*}_7$ sector of the model, so in
the whole $\Delta^p_5$-$\Delta^{p*}_5$, $\Delta^p_7$-$\Delta^{p*}_7$
sector of the NJL model which is under the color neutrality
constraint there are four massive pseudoscalar excitations: two of
them form an $\rm SU_c(2)$-antidoublet with mass $M^p_{D1}$, another
two particles form an $\rm SU_c(2)$-antidoublet with mass $M^p_{D2}$.

Now let us consider the 2SC ground state excitations in the
$\Delta^p_2$-$\Delta^{p*}_2$ sector. In this case, starting from
the effective action (\ref{3.380}), it is possible to obtain the
inverse propagator matrix $\Gamma^{2p}$ which is defined by the
relation (\ref{3.43}). In the rest frame, where $p=(p_0,0,0,0)$, its
Fourier-transformed matrix elements look like
\begin{eqnarray}
&&\Gamma^{2p}_{\Delta^{p*}_2\Delta^p_2}(p_0)=\frac{1}
{4H_p}-4i\int\frac{d^4q}{(2\pi)^4}
\left\{\frac{2m^2}{E^2}\cdot\frac{p_0+q_0-E^+}{(p_0+q_0)^2-
(E_\Delta^+)^2}\cdot\frac{q_0-E^-}{q_0^2-(E_\Delta^-)^2}
+\right.
\nonumber\\
&&~~~~~\left.+\frac{\vec q^2}{E^2}\Big [\frac{p_0+q_0-
E^+}{(p_0+q_0)^2-
(E_\Delta^+)^2}\cdot\frac{q_0+E^+}{q_0^2-(E_\Delta^+)^2}+
\frac{p_0+q_0+E^-}{(p_0+q_0)^2-
(E_\Delta^-)^2}\cdot\frac{q_0-E^-}{q_0^2-(E_\Delta^-)^2}\Big
]\right\},
\label{3.81}\\
&&\Gamma^{2p}_{\Delta^p_2\Delta^{p*}_2}(p_0)=
\Gamma^{2p}_{\Delta^{p*}_2\Delta^p_2}(-p_0),~~~~~~~~~~~~
\Gamma^{2p}_{\Delta^p_2\Delta^p_2}(p_0)=
\Gamma^{2p}_{\Delta^{p*}_2\Delta^{p*}_2}(p_0)=-4\Delta^2
\mathbb P(p_0), \nonumber\\
&&\mathbb P(p_0)=\int\frac{d^4q}{i(2\pi)^4}
\left\{\frac{2m^2}{E^2}\cdot\frac{1}{(p_0+q_0)^2-
(E_\Delta^+)^2}\cdot\frac{1}{q_0^2-(E_\Delta^-)^2}
+\right.
\nonumber\\
&&\left.+\frac{\vec q^2}{E^2}\Big [\frac{1}{(p_0+q_0)^2-
(E_\Delta^+)^2}\cdot\frac{1}{q_0^2-(E_\Delta^+)^2}
+\frac{1}{(p_0+q_0)^2-
(E_\Delta^-)^2}\cdot\frac{1}{q_0^2-(E_\Delta^-)^2}\Big ]\right\}.
\label{3.82}
\end{eqnarray}
Using in (\ref{3.81})-(\ref{3.82}) the substitution $\frac
1{4H_p}=\frac 1{4H_s}+\eta$ and then eliminating the coupling
constant $H_s$ in favor of another model parameters (with the help of
the gap equation (\ref{33})), we obtain after $q_0$-integrations:
\begin{eqnarray}
\label{3.83}
&&\mathbb P(p_0)=I_0(p_0^2)+m^2A(p_0),~~~~~
\Gamma^{2p}_{\Delta^{p*}_2\Delta^p_2}(p_0)=\eta +
(4\Delta^2-2p_0^2)I_0(p_0^2)+4p_0I_1(p_0^2)+m^2B(p_0),
\end{eqnarray}
where $I_0(p_0^2)$ and $I_1(p_0^2)$ are presented in (\ref{25}) and
(\ref{26}), respectively. The last terms in each of expressions
(\ref{3.83}) are proportional to $m^2$. In the 2SC phase the
constituent quark mass $m$ is a vanishingly small quantity (or it is
exactly zero if the current quark mass vanishes, $m_0=0$) as compared
with $\Delta$, etc (see Fig. 1), so we will ignore the contributions
of these terms in the  matrix elements (\ref{3.81})-(\ref{3.82}).
Thus, there is no need to have explicit expressions for the functions
$A(p_0)$ and $B(p_0)$ from (\ref{3.83}). In this approximation the
determinant of the inverse propagator matrix $\Gamma^{2p}(p_0)$ looks
like:
\begin{eqnarray}
&&{\rm det}\Gamma^{2p}(p_0)=
\Gamma^{2p}_{\Delta^{p*}_2\Delta^p_2}(p_0)
\Gamma^{2p}_{\Delta^p_2\Delta^{p*}_2}(p_0)-
\Gamma^{2p}_{\Delta^p_2\Delta^p_2}(p_0)
\Gamma^{2p}_{\Delta^{p*}_2\Delta^{p*}_2}(p_0)=
\nonumber\\
&&=\Big [\eta-2p_0^2I_0(p_0^2)\Big ]\Big
[\eta-2(p_0^2-4\Delta^2)I_0(p_0^2)\Big ]-16p_0^2I^2_1(p_0^2).
\label{3.89}
\end{eqnarray}
Obviously, at $\eta =0$ the equation ${\rm det}\overline
{\Gamma^{2p}} (p_0)=0$ coincides with eq. (\ref{032}) and has the
same solutions. The first one, $p_0^2=0$, corresponds to a stable
massless pseudo-scalar excitation, the second one lies in the second
Riemann sheet of the variable $p_0^2$. So it is a heavy pseudo-scalar
resonance and its mass and width are represented in the Fig. 4.
At small nonzero values of $\eta$ it is reasonable to suppose that
the equation ${\rm det}\overline {\Gamma^{2p}} (p_0)=0$ has a root
lying on the second Riemann sheet of $p_0^2$ as well. It might be
considered as a weak disturbance of the resonance solution of this
equation at $\eta =0$, so its mass and width behavior vs $\mu$ are
qualitatively the same as on the Fig. 4. Another solution of this
equation, $p_0^2=(M_{D3}^p)^2$, should not be significantly different
from the solution $p_0^2=0$ at $\eta =0$.  So, in searching of
$(M_{D3}^p)^2$, one can expand the expression (\ref{3.89})
into a Taylor-series of $p_0^2=0$:
\begin{eqnarray}
&&(M_{D3}^p)^2=\frac{\eta(\eta+8\Delta^2a)}
{16b^2+2a(\eta+8\Delta^2a)+2\eta(a-4\Delta^2a')},
\label{3.90}
\end{eqnarray}
where $a=I_0(0)$, $b=I_1(0)$, $a'=I_0'(0)$. Note that both the heavy
resonance and stable excitation with mass squared (\ref{3.90}) in the
pseudo-scalar diquark channel are singlets with respect to $\rm
SU_c(2)$. Since $a>0$, the expression (\ref{3.90}) is a positive one
at rather small and positive values of $\eta$. However, at
sufficiently small, but negative values of $\eta$, it is a negative
quantity, i.e. a tachyonic pseudo-diquark excitation appeared in the
model. This fact indicates the instability of the 2SC ground state
with $\vev {\Delta^{s}_{2}(x)}=\Delta$, $\vev {\Delta^{s}_{5}(x)}=0$,
$\vev {\Delta^{s}_{7}(x)}=0$, and $\vev {\Delta^{p}_{A}(x)}=0$.
Perhaps, in this case, i.e. at $H_p>H_s$, the phase with nonzero
ground state expectation values of pseudo-scalar diquarks, $\vev
{\Delta^{p}_{A}(x)}\ne 0$, should be realized.

As a result, we have shown that in the 2SC phase of the NJL model
(\ref{3}) there are five stable diquark excitations in the
pseudo-scalar channel. They form a singlet as well as two
antidoublets (in the case $H_p<H_s$) of the $\rm SU_c(2)$ group with
masses presented in (\ref{3.90}) and (\ref{3.70}), correspondingly.
Moreover, there is also a heavy resonance that is
$\rm SU_c(2)$-singlet with mass about 1100 MeV in this channel.

\subsection{The case of normal phase ($\Delta=0$,
$\mu_8=0$, $m\ne 0$)}\label{VIB}

Suppose that we are in the $\rm SU_c(3)$-symmetric (normal) phase of
our model, where $\Delta=0$. As it is easily seen from the color
neutrality constraint (\ref{35}), in this phase $\mu_8=0$. Moreover,
here $\breve E^\pm=E^\pm=E\pm\mu >0$, since in this phase $m>\mu$
(see Fig. 1 at $\mu<\mu_c$). Then, without loss of generality, it is
sufficient to study the mass spectrum, e.g., in the sector of
$\Delta^p_5$-$\Delta^{p*}_5$ diquarks. For the normal phase we have
from (\ref{3.66}):
\begin{eqnarray}
&&\Gamma^{5p}_{\Delta^{p*}_5\Delta^p_5}(p_0)=
\frac{1}{4H_p}-
16\int\frac{d^3q}{(2\pi)^3}
\frac{\vec q^2}{E}\frac{1}{4E^2-(p_0+2\mu)^2}\equiv \frac{1}{4H_p}-
F_p(\epsilon),
\label{3.71}
\end{eqnarray}
where $\epsilon =(p_0+2\mu)^2$ and the function $F_p(\epsilon)$ is
increasing on the interval $(-\infty,4m^2)$. Moreover, it is
analytical in the whole complex $\epsilon$-plane, except for the cut
$4m^2<\epsilon$ along the real axis. The masses of
pseudo-scalar diquarks are defined by the equation
\begin{eqnarray}
{\rm det}\Gamma^{5p}(p_0)
=\Gamma^{5ps}_{\Delta^{p*}_5\Delta^p_5}(p_0)
\Gamma^{5p}_{\Delta^p_5\Delta^{p*}_5}(p_0)=
\Gamma^{5p}_{\Delta^{p*}_5\Delta^p_5}(p_0)
\Gamma^{5p}_{\Delta^{p*}_5\Delta^p_5}(-p_0)=0,
\label{det3}
\end{eqnarray}
i.e. by zeros $\epsilon_0$ of the matrix element (\ref{3.71}) such
that $0<\epsilon_0<4m^2$ or lying in the second Riemann sheet of the
complex variable $\epsilon$. (The first one correspond to masses of
stable excitations, the second one -- to masses of resonances.) In
the present consideration we restrict ourselves to looking only for
stable pseudo-scalar diquarks. It is clear that there is
a single zero $\epsilon_0$ of (\ref{3.71}), obeying the condition
$0<\epsilon_0 <4m^2$, if and only if the coupling constant $H_p$ is
constrained by the relation $H^*_p<H_p<H_p^{**}$, where
\begin{eqnarray}
&&H^*_{p}=\frac {1}{4F_p(4m^2)}=\frac {\pi^2}{4\left
[\Lambda\sqrt{m^2+\Lambda^2}-m^2\ln((
\Lambda+\sqrt{m^2+\Lambda^2})/m)\right ]},\nonumber\\
&&H^{**}_{p}=\frac {1}{4F_p(0)}=\frac
{\pi^2\Lambda\sqrt{m^2+\Lambda^2}}{4\left
[3m^2\Lambda^2+\Lambda^4-3m^2\Lambda\sqrt{m^2+\Lambda^2}\ln((
\Lambda+\sqrt{m^2+\Lambda^2})/m)\right ]}.
\label{gamma3p}
\end{eqnarray}
(Note, $H^*_{p}=H^{**}_{s}$ from (\ref{gamma4}).) In this case the
masses of stable pseudo-scalar diquarks and antidiquarks are the
following
\begin{eqnarray}
(M^p_{\Delta})^2=(\sqrt{\epsilon_0}-2\mu)^2,~~~~~~
(M^p_{\Delta^*})^2=(\sqrt{\epsilon_0}+2\mu)^2,
\label{gamma6p}
\end{eqnarray}
respectively. (The mass splitting in (\ref{gamma6p}) is again
explained by the absence of a charge conjugation symmetry in the
presence of a chemical potential.) It follows from the underling
color $\rm SU(3)_c$ symmetry of the normal phase that at
$H^*_p<H_p<H_p^{**}$ there is indeed a color antitriplet of
pseudo-scalar diquarks  with the mass $M^p_{\Delta}$ as well as a
color triplet of pseudo-scalar antidiquarks with the mass
$M^p_{\Delta^{*}}$ (see (\ref{gamma6p})). For other regions of the
$H_p$-values, stable pseudo-scalar diquark excitations
of the $\rm SU_c(3)$-color symmetric ground state are forbidden.
 For a rather weak interaction in this channel ($H_p<H_p^*$),
 $\epsilon_0$ runs onto the second
Riemann sheet, and unstable pseudo-scalar diquark modes (resonances)
appear. Unlike this, a sufficiently strong interaction in this
channel ($H_p>H_p^{**}$) pushes $\epsilon_0$
towards the negative semi-axis, \textit{i.~e.\/}\ $(p_0+2\mu)^2<0$.
The later indicates a tachyonic singularity in the pseudo-scalar
diquark propagator, evidencing that the $\rm SU(3)_c$ color symmetric
ground state is not stable. In this case we guess that a
parity-breaking color superconducting phase is realized in which
the ground state expectation values of pseudo-scalar diquarks are not
zero, $\vev {\Delta^{p}_{A}(x)}\ne 0$.

Finally, a few words about diquarks in the particular case (recall,
$\mu<\mu_c$), $H_s=H_p\equiv H$, which corresponds to a NJL model
inspired by a one-gluon exchange approximation in QCD. In this case
we see that if $0<H<H_s^*$, then both scalar- and pseudo-scalar
diquarks are resonances. If $H_s^*<H<H_s^{**}$, then in the normal
phase, including the case $\mu=0$, only scalar diquarks are stable,
but pseudo-scalar ones are unstable particles. For larger values of
$H$ the normal phase is unstable in itself, since either or both
scalar diquark propagator (at $H_s^{**}=H_p^*<H<H_p^{**}$) and
pseudo-scalar one (at $H_p^{**}<H$) have tachyonic singularities.
Hence, one can conclude that at $H_s=H_p\equiv H$ scalar diquarks are
allowed to exist, at $H_s^*<H<H_s^{**}$, as stable excitations of the
normal phase. Pseudo-scalar diquarks in this phase are always
unstable particles (resonances).

\section{Summary and discussion}

In our previous papers \cite{bekvy,eky}, the masses of mesons and
diquarks, surrounded by moderately dense quark matter, were
investigated in the framework of NJL model (\ref{1}) at $H_p=0$, and
the color neutrality constraint was missed, for simplicity.
In the present paper, we have calculated the mass spectrum of meson
and diquark excitations in the color neutral cold dense quark matter.
We started from a low-energy  Nambu--Jona-Lasinio type effective
model (\ref{3}) for quarks of two flavors, with a quark chemical
potential $\mu$ and extended by including the chemical potential
$\mu_8$ of the 8th color charge. Moreover, the interaction in the
pseudo-scalar diquark channel was taken into account, in addition.
We considered only the interplay between normal- and 2SC phases.
This is a quite reasonable assumption in the framework of the model
(\ref{1}). Then, it was shown that in the presence of
color neutrality the transition to the 2SC phase occurred at a
somewhat smaller value of the quark chemical potential ($\mu_c=342$
MeV) than without this constraint ($\mu_c=350$ MeV).

It was proved in the present paper that in both models (\ref{1}) and
(\ref{3}), i.e. with or without a color neutrality constraint,
the $\sigma$-meson is mixed with the scalar diquark $\Delta_2^s$ in
the 2SC phase. In the previous paper \cite{eky} this mixing was
ignored in the consideration of the $\sigma$-meson mass.
At first, we have found that, if $\sigma$-$\Delta_2^s$ mixing is
ignored as in \cite{eky}, then the color neutrality requirement does
not change qualitatively the properties of $\pi$- and
$\sigma$-mesons, obtained in the framework of NJL model (\ref{1})
without the $\mu_8$ term. This is an expected result, since both
models (\ref{1}) and (\ref{3}) have an identical chiral symmetry.
Hence, at small values of $\mu_8$ (see Fig. 2) the meson masses
aquire small corrections as well (compare Fig. 2 from \cite{eky} and
Fig. 3 of our present paper). It follows from our consideration that
in this case (without mixing) both $\sigma$- and $\pi$-meson are
stable particles in the 2SC phase with masses of about 340 Mev (Fig.
3). However, if mixing is taken into account, then in the 2SC phase
the $\sigma$-meson is a resonance, decaying into a pair of quarks
with a rather small width $\sim$ 30 MeV (see Appendix). As far as we
know, the properties of $\pi$- and $\sigma$-mesons in the 2SC phase
have not been discussed in the literature before.

Moreover, the properties of scalar diquarks in the 2SC phase are
changed drastically, when the color neutrality condition is imposed.
Indeed, for the model (\ref{1}) we have found in the 2SC phase an
anomalous number of three Nambu-Goldstone (NG) bosons, the
$\rm SU_c(2)$-antidoublet of light diquarks, and a heavy resonance
that is an $\rm SU_c(2)$-singlet \cite{bekvy,eky}. Contrary, our
present investigation shows that in the model (\ref{3}) the scalar
diquark sector of the 2SC phase contains two real $\rm SU_c(2)
$-antidoublets of light excitations with the same mass $3|\mu_8|$,
one Nambu-Goldstone boson, and a heavy resonance with mass about 1100
MeV (see Fig. 4). To understand such a sharp difference in scalar
diquark masses, predicted by these two models, it is necessary to
compare their color symmetries. The first model, Lagrangian
(\ref{1}), is invariant under $\rm SU_c(3)$. However, in the second
model (\ref{3}) this symmetry is broken {\it explicitly}, due to the
presence of the $\mu_8$-term, to the subgroup $\rm SU_c(2)\times
U_{\lambda_8} (1)$. Then, in the 2SC phase, where the ground state is
an $\rm SU_c(2)$ invariant for both models (here
$\vev{\Delta^{s}_{2}(x)} \ne 0$, $\vev{\Delta^{s}_{5,7}(x)} =0$), we
have spontaneous breaking of the above mentioned symmetries. As a
consequence, there are five broken symmetry generators and an
abnormal number of three NG bosons for the model (\ref{1}) (the
explanation of this fact is presented in details in \cite{bekvy}). On
the other hand, for the model (\ref{3}) we have in the 2SC phase only
one broken $U_{\lambda_8}$-symmetry generator, resulting in a single
NG boson.

The properties of the pseudo-scalar diquarks in the 2SC phase depend
essentially on the relation between coupling constants $H_s$ and
$H_p$. First, note that at $H_p>H_s$ the 2SC phase is an unstable one
due to a negative mass squared (\ref{3.90}) of an $\rm
SU_c(2)$-singlet pseudo-scalar mode  (tachyonic instability). At
$H_p<H_s$ the pseudo-scalar excitations of this channel form in the
2SC phase two real stable $\rm SU_c(2)$-antidoublets with different
masses (\ref{3.70}) as well as the stable light $\rm SU_c(2)$-singlet
with mass (\ref{3.90}), and a heavy resonance with mass $\sim$ 1100
MeV.

We have also found that the antidiquark masses exceed those of the
diquarks in the normal $\rm SU_c(3)$ symmetric phase  (for $\mu
<\mu_c\approx 342$~MeV). This splitting of the masses is explained by
the violation of $C$-parity (charge conjugation) in the presence of a
chemical potential. In contrast, at $\mu =0$ the model is
$C$-invariant and all diquarks and antidiquarks of the same parity
have an identical mass. It follows from our investigation that stable
quark pair formation occurs in the scalar channel at a  weaker
coupling strength than in the pseudo-scalar one. Indeed,  if the
parameter set of the model is fixed by the relations (\ref{350}),
then in the normal phase scalar diquarks are stable particles (with
masses  about 700 MeV at $\mu=0$), since in this case $H_s^*
<H_s=3G/4 <H_s^{**}$ (see (\ref{gamma4})). However, only at
sufficiently high values of $H_p$, i.e. at $H_s^{**}=H_p^*<H_p
<H_p^{**}$ (see (\ref{gamma3p})), stable pseudo-scalar diquarks might
exist in the normal phase. On the other hand, if $H_p< H_s=3G/4$,
then in the normal phase pseudo-scalar diquarks are resonances,
decaying into two quarks. If $\mu$ exceeds the critical value $\mu_c$
and the system passes to the 2SC phase at $H_p<H_s=3G/4$, then five
of six pseudo-scalar diquarks aquire stability. Really, in
the 2SC phase, in contrast to the normal one, all particles move
inside the medium, so their decay might be prohibited by a Pauli
blocking principle (Mott effect). Suppose that $H_p=H_s\equiv H$ (in
this case the NJL model is inspired by a one-gluon exchange
approximation in QCD). Then at $H<H_s^{**}$ the normal phase is a
stable one, but at $H>H_s^{**}$ a tachyonic instability appears, so
the normal phase is destroyed (see the section \ref{VB}). We have
shown for this particular case that in the normal phase there might
exist stable scalar diquarks, however pseudo-scalar diquark modes are
always unstable excitations.

Of course, all observable particles render themselves as
colorless objects in the hadronic phase, and the diquarks are
expected to be confined, as they are no $\rm SU(3)_c$ color singlets.
Nevertheless, one may look at our and other related results on
diquark masses as an indication of the existence of  rather strong
quark-quark correlations inside baryons, which might help to explain
baryon dynamics. Some lattice simulations reveal  strong attraction
in the diquark channel \cite{wk} with a diquark mass $\sim$600~MeV.
Recently, in \cite{jw}, the mass and extremely narrow width, as
well as other  properties, of the pentaquark  $\Theta^+$  were
explained just on the assumption that it is composed of an antiquark
and two highly correlated $ud$-pairs. At present time, the nature of
the mechanism which may entail strong attraction of quarks in
diquark channels is actively discussed both in the nonperturbative
QCD and in other models (see, e.g., \cite{ripka} and references
therein).

For simplicity, we have studied the masses of the one-particle
excitations of the color neutral quark matter. Similar investigations
can be performed both in the color neutral and $\beta$-equilibrated
two-flavor NJL model, where electric charge neutrality is required in
addition, so one more (electric charge) chemical potential $\mu_Q$
must be introduced. In this case, for some range of the coupling
constant $H_s$  the gapless 2SC phase (g2SC) is realized (see, e.g.,
\cite{hs}, where the particular case $H_p=0$ of the NJL model
(\ref{1}) was considered). In contrast to the ordinary 2SC phase, two
additional gapless quark excitations then appeared in the g2SC phase,
but its meson and diquark mass spectrum, presumably, will not changed
qualitatively. Our belief is based on the structure of the ground
state expectation values of scalar diquarks in the g2SC phase, i.e.
$\vev{\Delta^{s}_{2}(x)} \equiv\Delta \ne 0$,
$\vev{\Delta^{s}_{5,7}(x)} =0$. This is expected to be identical to
that of our present consideration when only the color neutrality of
the quark matter is taken into account.

\section*{Acknowledgments}

We are grateful to D. Blaschke and M. K. Volkov for stimulating
discussions and critical remarks. After having completed the main
part of this work, we were informed by L. He, M. Jin, and P. Zhuang
on their paper \cite{end}. These authors get similar results for
light scalar diquark bosons in the case of color neutrality. We thank
them for informing us on their results. This work has been supported
in part by DFG-project 436 RUS 113/477/0-2, RFBR grant No.
05-02-16699, the Heisenberg--Landau Program 2004, and the ``Dynasty''
Foundation.

\appendix

\section{Correction to the $\sigma$-meson mass}

In  section \ref{IV} the numerical values for the $\sigma$-meson
mass $M_\sigma$ was obtained in the assumption that the mixing of
$\sigma$ with $\Delta_2^s$ is absent. Moreover, we have shown in this
approach that the $\sigma$-meson is a stable particle in the 2SC
phase. Now let us prove that a deeper investigation of the
$\sigma$-meson mass, based on the inclusion of the
mixing between $\sigma$ and $\Delta_2^s$, results in the conclusion
that in the 2SC phase $\sigma$ is a resonance, having a small decay
width into a quark pair.

Indeed, starting from the total number of effective actions
(\ref{2.24}), (\ref{40}), and (\ref{3.38}) it is possible to find the
3$\times$3 inverse propagator matrix ${\cal G}^{-1}$ of $\sigma$-,
$\Delta_2^s$-, and $\Delta_2^{s*}$ fields. In the center of mass
frame of the momentum space representation ($\vec p=0$) it has the
following form in the 2SC phase:
\begin{eqnarray}
{\cal G}^{-1}(p_0)=\left(\begin{array}[c]{lll}
\Gamma(p_0)~, & \Gamma_{\sigma\Delta^s_2}(p_0),&
\Gamma_{\sigma\Delta^{s*}_2}(p_0)\\
\Gamma_{\Delta^s_2\sigma}(p_0),&
\Gamma_{\Delta^s_2\Delta^s_2}^{2s}(p_0),&
\Gamma_{\Delta^s_2\Delta^{s*}_2}^{2s}(p_0)\\
\Gamma_{\Delta^{s*}_2\sigma}(p_0),&
\Gamma_{\Delta^{s*}_2\Delta^s_2}^{2s}(p_0),&
\Gamma_{\Delta^{s*}_2\Delta^{s*}_2}^{2s}(p_0)\end{array}\right),
\label{2100}
\end{eqnarray}
where
\begin{eqnarray}
\Gamma_{\sigma\Delta^s_2}(p_0)&=&\Gamma_{\Delta^{s*}_2\sigma}(p_0)=
\Gamma_{\sigma\Delta^{s*}_2}(-p_0)=\Gamma_{\Delta^s_2\sigma}(-p_0)=
\nonumber\\
&=&4m\Delta\int\frac{d^3q}{(2\pi)^3}
\left\{\frac{2E^++p_0}{EE_\Delta^+[4(E_\Delta^+)^2-p_0^2]}
+\frac{2E^--p_0}{EE_\Delta^-[4(E_\Delta^-)^2-p_0^2]}\right\}.
\label{230}
\end{eqnarray}
Other matrix elements from (\ref{2100}) are presented in
(\ref{22}) and (\ref{24}). Note that the matrix elements (\ref{230}),
mixing $\sigma$ with $\Delta_2^s$ and $\Delta_2^{s*}$, are
proportional to the dynamical quark mass $m$. So, in the 2SC phase
both $m$ and these matrix elements may be considered as  small
quantities (see Fig. 1). The mass spectrum is defined by the
equation ${\rm det}({\cal G}^{-1}(p_0))=0$, which has a rather
complicated form (note, ${\rm det}({\cal G}^{-1}(p_0))$ is an even
function vs $p_0$, i.e. it depends on $p_0^2$):
\begin{eqnarray}
&&F(p_0^2)\Gamma_{0}(p_0^2)=m^2\Delta^2f(p_0^2),
\label{210}
\end{eqnarray}
where
\begin{eqnarray}
&&F(p_0^2)=\Gamma^{2s}_{\Delta^s_2\Delta^s_2}
\Gamma^{2s}_{\Delta^{s*}_2\Delta^{s*}_2}-
\Gamma^{2s}_{\Delta^{s*}_2\Delta^s_2}
\Gamma^{2s}_{\Delta^s_2\Delta^{s*}_2}~~,~~~~~
m^2\Delta^2f(p_0^2)=
\Gamma_{\Delta_2\sigma}\Gamma_{\sigma\Delta_2
}\Gamma^{2s}_{\Delta^*_2\Delta^*_2}+
\Gamma_{\sigma\Delta^*_2}\Gamma_{\Delta^*_2\sigma}
\Gamma^{2s}_{\Delta_2\Delta_2}-\nonumber\\
&&~~~~~~~~~~~~~~~~-\Gamma_{\sigma\Delta_2}\Gamma_{\Delta^*_2\sigma}
\Gamma^{2s}_{\Delta_2\Delta^*_2}-
\Gamma_{\sigma\Delta^*_2}\Gamma_{\Delta_2\sigma}
\Gamma^{2s}_{\Delta^*_2\Delta_2}-
F(p_0^2)\Gamma_{1}(p_0^2),
\label{21000}
\end{eqnarray}
and the functions $\Gamma_{0}$, $\Gamma_{1}$ are defined in
(\ref{220}), (\ref{221}). In the $p_0^2$-plane the equation
(\ref{210}) has three solutions. One of them corresponds to a
Nambu-Goldstone boson, $p_0^2=0$. The other two we denote as
$(p_0^2)_\Delta$ and $(p_0^2)_\sigma$. Since $m$ is a small quantity,
the right hand side (RHS) of (\ref{210}) can be considered as a small
perturbation. Then the solution $(p_0^2)_\sigma$ (as well as
$(p_0^2)_\Delta$) can be constructed in the framework of a
perturbative expansion, based on the smallness of the RHS of
(\ref{210}). In the zeroth order we have
$F(p_0^2)\Gamma_{0}(p_0^2)=0$, so $(p_0^2)^{(0)}_\sigma=M_\sigma^2$,
where $M_\sigma$ is the zero of the function $\Gamma_{0}(p_0^2)$ and
is graphically represented in Fig. 3. One can easily obtain the first
perturbative correction
\begin{eqnarray}
(p_0^2)_\sigma=M_\sigma^2+\frac{m^2\Delta^2 f(M_\sigma^2)}
{F(M_\sigma^2)\Gamma^{'}_0(M_\sigma^2)}+
\cdots
\label{290}
\end{eqnarray}
and so on. Now let us put our attention on the fact that $f(p_0^2)$
and $F(p_0^2)$ are analytical functions in the whole complex
$p_0^2$-plane except the cut, composed from all real points such that
$4\Delta^2<p_0^2<\infty$. (In the rest of the real $p_0^2$-axis
$f(p_0^2)$ and $F(p_0^2)$ take real values.) Through the cut these
functions can be analytically continued to the second Riemann sheet.
In our case $M_\sigma\approx 350$ MeV and $\Delta\sim 100$ MeV,
so $M_\sigma^2\in (4\Delta^2,\infty)$, i.e. it lies on the cut.
Hence, both $f(M_\sigma^2)$ and $f(M_\sigma^2)$ have imaginary parts,
and the solution (\ref{290}) lies in the second Riemann sheet. It
means that $(p_0^2)_\sigma$ corresponds to a resonance in the mass
spectrum. Numerical estimates show that the width of this resonance
is a rather small quantity, less than 30 MeV.

Similar corrections can be easily performed for the heavy scalar
diquark resonance mass, corresponding to another solution
$(p_0^2)_\Delta$ of the equation (\ref{210}).

\end{document}